\documentstyle[12pt,epsbox]{article}
\newcommand {\yd}{\mbox{$\searrow$ \hspace{-7mm} $\swarrow$}}
\newcommand {\yl}{\mbox{$\swarrow$ \hspace{-7mm} $\nwarrow$}}
\newcommand {\yu}{\mbox{$\nearrow$ \hspace{-7mm} $\nwarrow$}}
\newcommand {\yr}{\mbox{$\nearrow$ \hspace{-7mm} $\searrow$}}
\newcommand {\yh}{\mbox{$\leftarrow$}}
\newcommand {\ym}{\mbox{$\rightarrow$}}
\newcommand {\msi}{\mbox{$\searrow$}}
\newcommand {\mue}{\mbox{$\nearrow$}}
\newcommand {\hsi}{\mbox{$\swarrow$}}
\newcommand {\hue}{\mbox{$\nwarrow$}}
\newcommand \sh[1]{{s}^{-}_{#1}}
\newcommand \st[1]{{s}^{+}_{#1}}
\newcommand \sha[1]{({s}^{-}_{#1})^{\ast}}
\newcommand \sta[1]{({s}^{+}_{#1})^{\ast}}
\newcommand \sa[1]{{{s}_{#1}}^{\ast}}

\newcommand {\del}{\partial}
\newcommand {\VERMA}{\cal V}
\newcommand {\DVERMA}{{\cal V}^{\ast}}
\newcommand {\FOCK}{\cal F}
\newcommand {\DFOCK}{{\cal F}^{\ast}}
\newcommand {\VERMAHQ}{{\cal V}(h,q)}
\newcommand {\DVERMAHQ}{{\cal V}^{\ast}(h,q)}

\newcommand \piint[1]{\oint \frac{#1}{2 \pi i}}
\newcommand \cm[2]{\left[ {#1}\, ,{#2} \right]}
\newcommand \ac[2]{\left\{ {#1}\, ,{#2}\right\}}
\newcommand \ket[1]{\left|{#1}\right\rangle}
\newcommand \bra[1]{\left\langle {#1}\right|}
\newcommand \bracket[2]{\left\langle{#1}|{#2}\right\rangle}
\newcommand {\bc}{\bf C}
\newcommand {\bz}{\bf Z}
\newcommand {\bm}{\bf M}
\newcommand {\bhz}{{\bf Z}+\frac{1}{2}}
\newcommand {\bphi}{\bar{\phi}}
\newcommand {\bpsi}{\bar{\psi}}
\newcommand {\ma}{\bar{a}}
\newcommand {\mb}{\bar{b}}
\newcommand {\aket}{| \alpha , \bar{\alpha} \rangle}
\newcommand {\abra}{\langle \alpha , \bar{\alpha} |}
\newcommand {\ld}{\cal L}
\newcommand \ya[1]{\stackrel{ Q^{(#1)}_{B}}{\leftarrow}}
\newcommand \ft[2]{{\cal C}^{#1}_{#2}}
\newcommand \vv[1]{{\bf V}_{#1}}
\newcommand {\bv}{\bf V}
\newcommand {\th}{{\cal H}^{+}}
\newcommand {\hh}{{\cal H}^{-}} 
\newcommand {\pmh}{{\cal H}^{\pm}}
\newcommand {\fprod}{\mathop{\prod}_{n=1}^{\infty}}
\newcommand \fsum[1]{\sum_{#1}^{\infty}}

\newcommand{\pr}{\hspace{\parindent}}

\begin{document}
\setlength{\oddsidemargin}{0cm}
\setlength{\baselineskip}{7mm}

\begin{titlepage}
    \begin{normalsize}
     \begin{flushright}
             {\bf UT-646} \\
             April 1993
     \end{flushright}
    \end{normalsize}
    \begin{LARGE}
       \vspace{1cm}
       \begin{center}
        {BRST analysis of N=2 superconformal \\
              minimal unitary models \\
          in Coulomb gas formalism } \\
       \end{center}
    \end{LARGE}

   \vspace{5mm}
\begin{center}
           Katsuyuki S{\sc ugiyama}
           \footnote{E-mail address:
              ksugi@tkyvax.phys.s.u-tokyo.ac.jp} \\
       \vspace{4mm}
                  {\it Department of Physics, University of Tokyo} \\
                  {\it Bunkyo-ku, Tokyo 113, Japan} \\
       \vspace{1cm}

     \begin{large} ABSTRACT \end{large}
\par
  \end{center}
\begin{quote}
\begin{normalsize}

We perform a BRST analysis of the N=2 superconformal minimal unitary models.
A bosonic as well as fermionic BRST operators are used to construct  
irreducible representations of the N=2 superconformal algebra 
on the Fock space as BRST cohomology classes of the 
BRST operators. Also a character formula is rederived 
by using the BRST analysis. 
 \end{normalsize}
\end{quote}

\end{titlepage}
\vfill

\section{Introduction}

\pr
N=2 superconformal field theories occupy fundamental positions
in the mathematical and theoretical physics.
First of all, the N=2 superconformal field theories 
provide an important class of 
topological field theories 
by means of the "twisting" operation \cite{twist}.
Perturbing these topological field theories, we obtain 
massive N=2 supersymmetric theories \cite{massive} 
which have pseudo-topological properties. 
Together with the non-renormalization theorems, it is even possible
to follow renormalization group flows exactly and 
obtain the various exact results.
$N=2$ supersymmetric theories are also very useful 
tools in describing the classical vacua 
of Calabi-Yau manifolds when we consider the string compactification 
with space-time supersymmetry. These theories 
appear in the mirror symmetry which make it possible to compute 
exactly some non-perturbative effects on the world sheet.

In view of the intense research activities in various areas 
of the N=2 supersymmetric theories, 
it seems important to consolidate the fundamental bases of 
these theories. 
In this paper we use the Coulomb gas formalism of the $N=2$ theories 
and investigate irreducible representation spaces 
of the N=2 minimal unitary models \cite{minimal,MSS,KM}. 
We discuss the structure of the $N=2$ Fock space and 
screening operators in detail and obtain a resolution 
of Felder type \cite{Fel} by considering cohomologies of screening operators. 
In this paper we use the standard Coulomb gas formalism \cite{KM,kito}, 
and obtain a result different 
from that of V. Sadov \cite{Sadov,DCG}, who analyzed the 
N=2 super Virasoro models 
by taking the Hamiltonian reduction.
\newline\\
This paper is organized as follows. In section 2, we review briefly  
the N=2 superconformal minimal unitary models. In section 3, 
we explain the N=2 Coulomb gas formalism, 
Fock spaces to fix the notations 
in this paper. We write down the bosonic, and fermionic
screening operators and specify their 
integration contours.
In section 4, we explain the null states, 
the Kac determinant formula \cite{KM,BFK}, and its factorization. 
From the factorized formula of the Kac determinant, 
we suggest the embedding patters of the Fock space submodules.  
In section 5, the actions of the BRST screening operators are studied, 
and the BRST cohomologies of the Fock space are investigated. 
A resolution of this Fock space is obtained by considering 
the relations between the Verma module, the dual Verma module, 
and the Fock module. As an application, we derive the character formula 
from the BRST analysis. In section 6 is devoted to conclusions and comments. 
In the Appendix, we write down a few null states at some lower levels. 
\\\\
\section{N=2 minimal unitary models}

\pr
This section is a short review of minimal unitary series of 
N=2 superconformal field theories. 
We present the commutation relations of the N=2 superconformal algebra. 
For details, we refer to the literatures \cite{minimal,MSS,KM}.
\subsection[N=2 minimal unitary models]{N=2 superconformal algebras} 

\pr
The N=2 superconformal field theories \cite{minimal,MSS,KM} 
are the extention of the 
Virasoro theories to include a complex $(N=2)$ supersymmetry on the 
Riemann surface.
Basic commutation relations are given by
\begin{eqnarray}
\cm{L_{n}}{L_{m}}&=&(n-m)L_{n+m}+\frac{c}{12}n(n^{2}-1)\delta_{n+m,0} 
       \mbox{\hspace{20mm} ,}       \label{eqn:LNLM}\\
\cm{L_{n}}{G^{+}_{r}}&=& 
    \left( \frac{n}{2}-r \right) G^{+}_{n+r}  \mbox{\hspace{58mm},}  \\
\cm{L_{n}}{G^{-}_{s}}&=& 
     \left( \frac{n}{2}-s \right) G^{-}_{n+s}  \mbox{\hspace{58mm},}  \\
\cm{L_{n}}{J_{m}}&=& -mJ_{n+m}  \mbox{\hspace{68mm},}  \\
\ac{G^{+}_{r}}{G^{-}_{s}}&=& L_{r+s}+\frac{1}{2}(r-s)J_{r+s}+\frac{c}{6} 
         \left( r^{2}-\frac{1}{4} \right) 
           \delta_{r+s,0}  \mbox{\hspace{10mm},}  \label{eqn:TGHG} \\
\cm{J_{n}}{G^{+}_{r}}&=& +G^{+}_{n+r}  \mbox{\hspace{71mm},}  \\
\cm{J_{n}}{G^{-}_{s}}&=& -G^{-}_{n+s}  \mbox{\hspace{71mm},}  \\
\cm{J_{n}}{J_{m}}&=& \frac{c}{3}n \delta_{n+m,0}  \mbox{\hspace{65mm},}  \\
\ac{G^{+}_{r}}{G^{+}_{r'}}&=&0  \mbox{\hspace{80mm},}  \\
\ac{G^{-}_{s}}{G^{-}_{s'}}&=&0  \mbox{\hspace{80mm}.}  \label{eqn:TGTG} 
\end{eqnarray}
where $L_{n}$ is the generator of the Virasoro algebra and $c$ is the central 
charge. ${G}^{+}_{r}$,${G}^{-}_{s}$ are coefficients of 
the mode expansions of superstress tensors. $J_{n}$ is the generator of 
the $U(1)$ symmetry. 
The subscripts of $L$ and $J$ are integers, and that of $G^{\pm}$ 
are integers modulo $\pm a$. The parameter $a$ reflects 
algebra automorphisms of the $N=2$ superconformal algebra called the 
"spectral flow". By the spectral flow, the Neveu-Shwarz sector 
can be transferred to the Ramond sector and vice versa. 
So we restrict our discussions 
to the NS sector without loss of generality. 
\subsection[N=2 minimal unitary models]{Verma module}

\pr
The Verma module ${\VERMA} (c;h,q)$ of 
the N=2 superconformal algebra is spanned by the heighest weight state 
$\ket{h,q}$ and the vectors,
\[
J_{-m_{1}}\cdots J_{-m_{r}}L_{-n_{1}}\cdots L_{-n_{s}}
G^{+}_{-k_{1}}\cdots G^{+}_{-k_{t}}G^{-}_{-l_{1}}
\cdots G^{-}_{-l_{u}}\ket{h,q} \,\,,
\]
which occur at a level, 
\[
N=\sum^{r}_{i=1}m_{i}+\sum^{s}_{i=1}n_{i}
 +\sum^{t}_{i=1}k_{i}+\sum^{u}_{i=1}l_{i}\,\,,
\]
and with a relative charge, 
\[
R=t-u \,\,.
\]
The Verma module $\VERMAHQ$ is symbolically expressed as,
\begin{eqnarray*}
{\cal V}(h,q) 
&=& \sum_{I,K,M,N}{\bf C}J_{-I}L_{-K}G^{+}_{-M}G^{-}_{-N}\ket{h,q} \,\,\,.
%&=& {\bigoplus}_{N\geq 0,\,R\geq 0}{\bf C}{\cal V}_{N,\,R}(h,q)
\end{eqnarray*}
%where ${\VERMA}_{N,\,R}(h,q)$ are submodules of ${\VERMA} (h,\, q)$ 
%with eigenvalues $(h+N,\,q+R)$.
%The lowest energy ground state is an $OSp(2|2)$ invariant vacuum $\ket{0}$ 
%which satisfies conditions.
%\begin{eqnarray*}
%L_{n}\ket{0} &=& 0 \mbox{\hspace{1cm}} (n\geq -1,\,\,n \in {\bf Z})\\
%J_{m}\ket{0} &=& 0 \mbox{\hspace{1cm}} (m\geq 0,\,\,m \in {\bf Z})\\
%G^{\pm}_{r}\ket{0} &=& 0 \mbox{\hspace{1cm}} 
%   (r\geq -\frac{1}{2},\,\,r \in \frac{1}{2}+{\bf Z})
%\end{eqnarray*}
%A primary field $\varphi (z)$ of the N=2 superconformal field 
%theories satisfies ;
%\begin{eqnarray*}
%&&T(z)\varphi (w) 
%       \sim \frac{h}{(z-w)^{2}}\varphi (w) +\frac{\del \varphi (w)}{z-w}\\
%&&J(z)\varphi (w) \sim \frac{q}{z-w}\varphi (w)\\
%&&G^{\pm}(z)\varphi (w) \sim \frac{1}{z-w} {\rho}^{\pm}(w)
%\end{eqnarray*}
%where ${\rho}^{\pm}$ are superpartners of $\varphi$. 
%Then the state $\ket{h,q}$ is created from the vacuum state $\ket{0}$. 
%\[
%\ket{h,q} \equiv \lim _{z \rightarrow 0}{\varphi}_{h,q}(z)\ket{0}
%\]

The dual Verma module $\DVERMAHQ$
is defined similarly. 
We may write the ${\DVERMA}$ as,
\[
\DVERMAHQ = \sum_{I,K,M,N} {\bf C} \bra{h,q}G^{-}_{N}G^{+}_{M}L_{K}J_{I} \,\,.
\]
As is well-known, the contravariant form $\langle |\rangle$ is 
not necessarily positive definite.
In order to make the form positive definite, the value of the central charge c 
should be restricted to a set of special values,
${c} \geq 3, \mbox{or \hspace{2mm}} 
c= 3 \left({1-\frac{2}{M}}\right)$ \hspace{2mm} $(M=2,3,4, \cdots)$
\cite{minimal,MSS,BFK}. 
We then
restrict the module by subtracting the submodules 
generated by zero norm states (null states). 
\subsection[N=2 minamal unitary models]{Unitarity condition}

\pr
Let us consider the unitarity condition. From the relation (\ref{eqn:TGHG}), 
the following relation is satisfied,
\begin{eqnarray*}
\bra{h,q} G^{+}_{r}G^{-}_{-r} \ket{h,q}
= h+rq+ \frac{c}{6} \left({r^{2}-\frac{1}{4}}\right) \geq 0 \,\,,
\end{eqnarray*}
for all positive half integers $r$. This condition can be rewritten as, 
\begin{eqnarray}
h \geq \frac{1}{2}|q| \label{eqn:unitary} \,\,.
\end{eqnarray}
The allowed values of $h$, $q$ and $c$ in the unitary 
irreducible representations 
are known \cite{MSS,Gep}; 
\begin{eqnarray*}
&&c=3 \left( {1-\frac{2}{M}} \right) \,\,, \\
&&h=\frac{l(l+2)-m^{2}}{4M}, q=\frac{m}{M}
\end{eqnarray*}
with $l=0,1,2,\cdots ,M-2$ and $m=-l,-l+2,\cdots ,l-2,l$ \hspace{2mm}
$(M=2,3,4,\cdots)$.
If the condition (\ref{eqn:unitary}) is not satisfied \hspace{3mm}
($i.e. h< \frac{1}{2}|q|$), \hspace{2mm}
$l< |m|$ leads to states whose magnitude of magnetic 
quantum number exceeds the value of spin. 
It is reasonable that such a case is 
excluded by imposing unitarity.
\section{N=2 Coulomb gas formula}

\pr
In this section, an N=2 Coulomb gas formula and its Fock space are introduced. 
Then BRST operators are constructed by integrating the screening operators
along appropriate contours. 
Their nilpotencies are proved. 
\subsection[N=2 Coulomb gas formula]{Coulomb gas formula}

\pr
The N=2 superconformal algebra is expressed by using 
a pair of complex scalar fields $\phi, \bphi$ and a pair of 
complex fermions $\psi, \bpsi$. The fields of each pair are complex conjugate 
of each other and are represented using mode expansions,  
\begin{eqnarray*}
&&\phi (z)=q-i(p-\beta)\log z +i\sum_{n\neq 0}\frac{1}{n}a_{n}z^{-n} \,\,,\\
&&\bphi (z)=\bar{q}-i(\bar{p}-\bar{\beta})\log z 
             +i\sum_{n\neq 0}\frac{1}{n} {\ma}_{n} z^{-n} \,\,,\\
&&\psi (z)=\sum_{r \in \bhz}b_{r}z^{-r-\frac{1}{2}} \,\,,\\
&&\bpsi (z)=\sum_{r \in \bhz}{\bar{b}}_{r}z^{-r-\frac{1}{2}} \,\,,
\end{eqnarray*}
where the shifts of the momenta of $\phi, \bphi$ 
$i.e. \mbox{\hspace{2mm}} \beta,\bar{\beta}$ are 
constants and come from the background charge of 
the Feigin-Fuchs formulation. 
These oscillators satisfy the commutation relations,
\begin{eqnarray*}
&&\cm{q}{{\bar{a}}_{0}}=i \mbox{\hspace{1cm}} , \,\,
 {\bar{a}}_{0} \equiv \bar{p}-\bar{\beta} \,\,,\\
&&\cm{\bar{q}}{a_{0}}=i \mbox{\hspace{1cm}} , \,\,
 {a}_{0} \equiv p-\beta \,\,,\\
&&\cm{a_{n}}{{\bar{a}}_{m}}=n {\delta}_{n+m,0} \,\,,\\
&&\left\{ {b_{r},\bar{b}_{s}}\right\}={\delta}_{r+s,0} \,\,.
\end{eqnarray*}
The central charge of this algebra is given by  
$c=3(1-2 {\beta}^{2})$. 
%We can write $N=2$ generators  
%by using mode expansion. 
%\begin{eqnarray*}
%L_{n} &=& \sum_{\stackrel{\scriptstyle l+m=n}{l,m \in \bz}}\mbox{:}
%        a_{l} {\ma}_{m} \mbox{:} +\frac{1}{2}(n+1)
%        (\beta {a}_{n} +\bar{\beta} {\ma}_{n})
%        +\frac{1}{2} \sum_{\stackrel{\scriptstyle r+s=n}{r,s \in \bhz}}
%        s \mbox{:} ( b_{r} {\mb}_{s} +{\mb}_{r} b_{s} ) \mbox{:}\\
%G^{+}_{r} &=& -i \sum_{\stackrel{\scriptstyle s+m=r}
%        {m \in {\bz} ,s \in {\bhz}}}
%           \mbox{:} b_{r} {\ma}_{m} \mbox{:} -i\beta 
%           \left({r+\frac{1}{2}}\right) b_{r} \\
%G^{-}_{r} &=& i\sum_{\stackrel{\scriptstyle s+m=r}{m \in {\bz} ,s \in {\bhz}}}
%           \mbox{:} {\mb}_{r} a_{m} \mbox{:} +i\bar{\beta} 
%           \left({r+\frac{1}{2}}\right) {\mb}_{r} \\
%J_{n} &=& \sum_{\stackrel{\scriptstyle s+r=n}{s,r \in \bhz}}
%         \mbox{:} b_{s} {\mb}_{r} \mbox{:} 
%           +(\beta a_{n} - \bar{\beta} {\ma}_{n}) 
%\end{eqnarray*}
To simplify the treatment, we put $\beta = \bar{\beta} 
 (i.e.\mbox{\hspace{2mm} real})$ in the rest of this paper. 
\subsection[N=2 Coulomb gas formula]{Fock space}

\pr
The Hilbert space of this model is a $"charged"$ Fock space. 
This space ${\cal F} (\alpha , \bar{\alpha})$ with central charge 
$c=3(1-2 \beta \bar{\beta})$ is defined as a representation space of 
the Heisenberg algebras of the oscillators.
The Fock module is constructed on a ground state $\ket{\alpha, \bar{\alpha}}$ 
which satisfies the relations, 
\begin{eqnarray*}
%&&{a}_{n} \ket{\alpha , \bar{\alpha}}=0 \\
%&&{\ma}_{n} \ket{\alpha , \bar{\alpha}}=0 \mbox{\hspace{2cm}} (n \geq 1) \\
&&{a}_{0} \ket{\alpha , \bar{\alpha}}= 
   \alpha \ket{\alpha , \bar{\alpha}} \,\,,\\
&&{\ma}_{0} \ket{\alpha , \bar{\alpha}}
    = \bar{\alpha} \ket{\alpha , \bar{\alpha}} \,\,, 
%&&{b}_{r} \ket{\alpha , \bar{\alpha}}=0 \\
%&&{\mb}_{r} \ket{\alpha , \bar{\alpha}}=0 \mbox{\hspace{2cm}} (r \geq 1/2) \\
\end{eqnarray*}
%and expanded by $\ket{\alpha , \bar{\alpha}}$ and 
%\[
%{a}_{{-n}_{1}} \cdots {a}_{{-n}_{i}} {\ma}_{{-m}_{1}} \cdots {\ma}_{{-m}_{j}}
%{b}_{{-r}_{1}} \cdots {b}_{{-r}_{k}} {\mb}_{{-s}_{1}} \cdots {\mb}_{{-s}_{l}}
%\ket{\alpha , \bar{\alpha}}
%\]
%The Fock module is 
and symbolically represented as, 
\[
{\cal F}(\alpha , \bar{\alpha})
  =\sum {\bf C} {a}_{-I} {\ma}_{-J} {b}_{-M} {\mb}_{-N} 
   \ket{\alpha , \bar{\alpha}}\,\,.
\]
One introduces the structure of an N=2 superconformal algebra 
(N=2 SCA) module on this Fock space by using the previous oscillator 
formalisms of the N=2 generators. 
%Especially the state $\ket{\alpha , \bar{\alpha}}$ is the eigenstate of 
%the operators $L_{0}, J_{0}$ . 
%\begin{eqnarray*}
%L_{0} \ket{\alpha , \bar{\alpha}}
%     &=& \left\{ {{\alpha} \bar{\alpha}
%      + \frac{1}{2} \beta ({\alpha}+{\bar{\alpha}})}\right\} 
%      \ket{\alpha , \bar{\alpha}} \\
%J_{0} \ket{\alpha , \bar{\alpha}} 
%     &=& \beta ({\alpha} -{\bar{\alpha}}) \ket{\alpha , \bar{\alpha}} \\
%\end{eqnarray*}
An N=2 module on the Fock space is represented,
\[
{\cal F}_{\alpha , \bar{\alpha}}= \sum {\bf C} J_{-I} L_{-K} 
        G^{+}_{-M} G^{-}_{-N} \ket{\alpha , \bar{\alpha}} \,\,,
\]
with the previous relations.
On the other hand, the dual module ${\DFOCK}_{\alpha , \bar{\alpha}}$ 
of ${\FOCK}_{\alpha , \bar{\alpha}}$ is defined,
\[
{\DFOCK}_{\alpha , \bar{\alpha}} \equiv \sum {\bf C} {\abra} 
      G^{-}_{N} G^{+}_{M} L_{K} J_{I} \,\,.
\]
%with the definitions. 
%\begin{eqnarray*}
%&&\abra {a}_{-n} =0 \\
%&&\abra {\ma}_{-n} =0 \hspace{2cm} (n>0) \\
%&&\abra {b}_{-r} =0 \\
%&&\abra {\mb}_{-r}=0 \hspace{2cm} (r \geq 1/2) \\
%&&\abra {a}_{0} =\abra \alpha \\
%&&\abra {\ma}_{0} =\abra \bar{\alpha}
%\end{eqnarray*}
%A structure of ${\DFOCK}_{\alpha , \bar{\alpha}}$ as an N=2 module is 
%specified to determine the action of the N=2 generators on a general 
%covector $\omega$. 
One defines the action of $N=2$ generators 
on the dual Fock space ${\DFOCK}_{\alpha , \bar{\alpha}}$,
%by using the contravariant form.
\begin{eqnarray*}
\omega \in \DFOCK \mbox{\hspace{1cm}} , \,\,  \xi \in \FOCK \,\,,\\
\bracket{L_{n} \omega}{\xi} &=& \bracket{\omega}{L_{-n} \xi} \,\,,\\
\bracket{G^{\pm}_{r} \omega}{\xi} &=& 
   \bracket{\omega}{G^{\mp}_{-r} \xi} \,\,,\\
\bracket{J_{n} \omega}{\xi} &=& \bracket{\omega}{J_{-n} \xi}\,\,.
\end{eqnarray*}
\subsection[N=2 Coulomb gas formula]{Vertex operators}

\pr
The vertex operators are defined \cite{MSS,KM,kito} by
\begin{eqnarray*}
V^{0}_{\alpha , \bar{\alpha}} &=& 
  \mbox{:} e^{i \alpha \bphi + i \bar{\alpha} \phi} \mbox{:} \,\,.
%V^{1}_{\alpha , \bar{\alpha}} &=& 
%  \mbox{:} i \bar{\alpha} \psi 
%   e^{i(\bar{\alpha \phi + \alpha \bphi})} \mbox{:} \\
%V^{1}_{\alpha , \bar{\alpha}} &=& 
%  \mbox{:} i \alpha \bpsi e^{i(\bar{\alpha} \phi + \alpha \bphi)} \mbox{:} \\
%V^{2}_{\alpha ,\bar{\alpha}} &=& 
%  \mbox{:} \left\{{\alpha \bar{\alpha} {\psi} {\bpsi} 
%  + \frac{1}{2} i \beta (\del \phi - \del \bphi)}\right\} 
%  e^{i(\bar{\alpha} \phi +\alpha \bphi)} \mbox{:} 
\end{eqnarray*}
%The conformal weight $h$ and the $U(1)$ charge $q$ of 
%$V^{0}_{\alpha , \bar{\alpha}}$ are 
%\begin{eqnarray*}
%h &=& {\alpha} {\bar{\alpha}} + \frac{1}{2} {\beta} 
%   \left({\alpha}+{\bar{\alpha}}\right) \\
%q &=& {\beta} \left({\alpha}-{\bar{\alpha}}\right)
%\end{eqnarray*}
%respectively as we claim. 
For convenience, we parametrize $(\alpha , \bar{\alpha})$ 
\cite{BFK,Matsu} by half-integers 
$(j ,k) \mbox{\hspace{3mm}} \left({j,k \in \bhz}\right) $,
\begin{eqnarray*}
{\beta}^{2} &=& \frac{1}{M} \,\,,\\
(\alpha , \bar{\alpha}) 
  &=& \left({\beta (j-\frac{1}{2}) , \beta (k-\frac{1}{2})}\right) \,\,,\\
V_{j,k} & \equiv & \mbox{:} e^{i \beta (j-\frac{1}{2}) \bphi 
  + i \beta (k-\frac{1}{2}) \phi} \mbox{:} \,\,,\\
h &=& \frac{1}{M} \left({jk-\frac{1}{4}}\right) \,\,,\\
q &=& \frac{1}{M}(j-k) \,\,,\\
c &=& 3 \left( {1-\frac{2}{M}} \right) \,\,.
\end{eqnarray*}
%Then vertex operators with $(h,q)=(1,0)$ are called screening operators. 
In our model, there are three kinds of 
screening operators with $(h,q)=(1,0)$.  
\begin{eqnarray*}
&&V_{B} \equiv \mbox{:} 
 \left\{ {{\beta}^{2} \psi \bpsi -\frac{1}{2} i 
 {\beta} (\del {\phi} - \del {\bphi})} \right\} 
  e^{-i {\beta} ({\phi} +{\bphi})} \mbox{:} \,\,,\\
&&V_{f} \equiv \mbox{:} 
 \frac{i}{\beta} \psi e^{i \frac{1}{\beta} \phi} \mbox{:} \,\,,\\
&&{\bar{V}}_{f} \equiv \mbox{:} 
  \frac{i}{\beta} \bpsi e^{i \frac{1}{\beta} \bphi} \mbox{:} \,\,.
\end{eqnarray*}
Using these screening operators, BRST operators are defined as,
\begin{eqnarray}
&&Q_{f} \equiv \piint{dz} \mbox{\hspace{2mm}} V_{f}(z) \,\,,\\
&&{\bar{Q}}_{f} \equiv \piint{dz} \mbox{\hspace{2mm}} {\bar{V}}_{f}(z) \,\,,\\
&&Q^{(n)}_{B} \equiv 
  \oint_{C_{1}} \frac{{dz}_{1}}{2 \pi i} 
  \oint_{C_{2}} \frac{{dz}_{2}}{2 \pi i}\cdots 
  \oint_{C_{n}} \frac{{dz}_{n}}{2 \pi i} \mbox{\hspace{2mm}} 
  V_{B}({z}_{1}) V_{B}({z}_{2}) \cdots V_{B}({z}_{n}) \,\,, \label{eqn:QB}
\end{eqnarray}
where $Q_{f}, {\bar{Q}}_{f}$ are fermionic BRST operators. On the other hand, 
$Q_{B}$ is a bosonic BRST operator of the Felder type. 
The fermionic BRST operators satisfy the following relations,
\begin{eqnarray*}
Q^{2}_{f}={\bar{Q}}^{2}_{f}=0\,\,, \\
\ac{Q_{f}}{{\bar{Q}}_{f}} =0 \,\,,\\
\cm{Q_{f}}{{Q}^{(n)}_{B}} = \cm{\bar{{Q}_{f}}}{{Q}^{(n)}_{B}} =0 \,\,.
\end{eqnarray*} 
In the Felder type BRST operator $Q_{B}$, integration contours 
have to be specified.
The choice of the contours 
$C_{1},C_{2}, \cdots, C_{n}$ is illustrated in the Figure 5 and 6.
%\vspace{5cm} 
We may rearrange the $V_{B}$'s in the integrand so that their arguments 
satisfy,
\[
\arg {z_{1}} < \arg {z_{2}} < \cdots < \arg {z_{n}}\,\,.
\]
Taking into account of all possible
orderings of the arguments, 
%in the formula $Q^{(n)}_{B}$, 
the multiple-integral can be rewritten,
\begin{eqnarray}
Q^{(n)}_{B} = \mathop{\prod}_{l=1}^{n} 
  \frac{e^{2 {\pi} i l {\beta}^{2}} -1}{e^{2 {\pi} i {\beta}^{2} } -1} 
   \begin{array}[t]{c}
    {\displaystyle  {\piint{d {z}_{1}} \piint{d {z}_{2}}} 
    \cdots \piint{d {z}_{n}} } \\
    {\scriptstyle 0 < \arg {{z}_{1}} < \arg {{z}_{2}} < 
        \cdots < \arg {{z}_{n}} < 2 \pi} 
  \end{array}
  \mbox{\hspace{2mm}}
   {V}_{B} ({z}_{1}) {V}_{B} ({z}_{2}) 
     \cdots {V}_{B} ({z}_{n}) \,\,. \label{eqn:BQ}
\end{eqnarray}
From this expression, we obtain the important result, 
\[
Q^{(n=M)}_{B} =0\,\,.
\]

Null states are constructed by applying these BRST operators on
primary vertex operators, and operators so obtained are called 
screened vertex operators \cite{KM,Fel}. 
Firstly, let us consider the case of fermionic BRST operators.
The screened vertex operator 
$Q_{f} V_{j,k}(z) = \piint{dw} V_{f}(w) V_{j,k}(z)$ is single-valued 
under rotation $(z,w) \rightarrow (e^{2 \pi i} z, e^{2 \pi i} w)$.
Also from $Q^{2}_{f} =0$, the BRST operator can act on $V_{j,k}$ only once.
For ${\bar{Q}}_{f} V_{j,k}(z)$, the similar relation can be given. 
The number $n$ of $V_{B}$'s in the bosonic BRST operator ${Q}^{(n)}_{B}$ is
determined from the single valuedness of ${Q}^{(n)}_{B} V_{j,k} $,
\[
\exp \left\{{2 \pi i \left[{\frac{1}{2} n(n-1) \cdot 2 {\beta}^{2}
  - \left({j-\frac{1}{2}}\right) {\beta}^{2} n
  - \left({k-\frac{1}{2}}\right) {\beta}^{2} n }\right]}\right\} =1 \,\,.
\]
Using ${\beta}^{2} =\frac{1}{M}$, we obtain a value $n$,  
\[
n=j+k+Mm \hspace{2cm}(m \in \bz) \,\,.
\]
Moreover because of $Q^{(M)}_{B} =0$, we can determine the integer 
$m$ uniquely for 
a given value of $(j,k)$ by imposing 
the inequality $ 0<n<M $. These BRST operators 
$Q_{f},{\bar{Q}}_{f}, \mbox{ and \hspace{1mm} } Q^{(n)}_{B}$ 
can be seen as mappings from a Fock module 
${\FOCK}_{j,k}$ (denoting ${\FOCK}_{\alpha , \bar{\alpha}}$ with 
$\alpha =\beta \left({j-\frac{1}{2}}\right) , 
\bar{\alpha} =\beta \left({k-\frac{1}{2}}\right)$ as ${\FOCK}_{j,k})$ 
to appropriate Fock modules,
\begin{eqnarray*}
Q_{f} & \mbox{:} & {\FOCK}_{j,k} \rightarrow {\FOCK}_{j,k+M} \,\,,\\
{\bar{Q}}_{f} & \mbox{:} & {\FOCK}_{j,k} \rightarrow {\FOCK}_{j+M,k} \,\,,\\
Q^{(n)}_{B} & \mbox{:} & {\FOCK}_{j,k} \rightarrow {\FOCK}_{j-n,k-n} \,\,.
\end{eqnarray*}
These actions of BRST operators allow us to 
introduce suitable gradings of Fock modules. 
\section{Null states}

\pr
In this section, null states are explained. 
Where the null states appear could be 
determined by making use of the Kac determinant \cite{KM,BFK}. 
By studying the determinant formula carefully, 
the relations between the Verma module and the Fock module 
or between the dual Verma module and the Fock module could be obtained. 
Then we suggest an embedding diagram of an N=2 superconformal module on the 
Fock space in the Coulomb gas formalism. 
\subsection[Null states]{Kac determinant}

\pr
A null state in the Verma module is defined as a state which is orthogonal 
to any other state in the dual Verma module. 
Especially the null state is orthogonal 
to itself, and is a zero norm state. 
%A null state $\ket{\chi}$ 
%in the Verma module of the N=2 superconformal algebra could be 
%defined. 
%\begin{eqnarray*}
%&&L_{n} \ket{\chi} =0 \hspace{2cm}(n \geq 1) \\
%&&J_{n} \ket{\chi} =0 \hspace{2cm}(n \geq 1) \\
%&&G^{\pm}_{r} \ket{\chi} =0 \hspace{2cm}(r \geq 1/2) \\
%&&\ket{\chi} =\sum C_{IKMN} J_{-I} L_{-K} G^{+}_{-M} G^{-}_{-N} \ket{primary} 
%\end{eqnarray*}
%A null state $\bra{\chi}$ in the dual Verma module is defined similarly.
%\begin{eqnarray*}
%&&\bra{\chi} L_{-n} =0 \hspace{2cm}(n \geq 1) \\
%&&\bra{\chi} J_{-n} =0 \hspace{2cm}(n \geq 1) \\
%&&\bra{\chi} G^{\pm}_{-r} =0 \hspace{2cm}(r \geq 1/2) \\
%&&\bra{\chi} = \sum C_{IKMN} \bra{primary} G^{-}_{N} G^{+}_{M} L_{K} J_{I}
%\end{eqnarray*}
%It is these zero norm properties of null states that prevent us 
%from obtaining irreducible representations of the N=2 superconformal algebra, 
%because the contravariant form is positive semi-definite owing to 
%these null states. 
We explicitly construct a few examples of null states at  lower levels 
in the Appendix. 
%In order to get the irreducible representations, 
%the information about the null states 
%(especially the levels and the relative charges at which they appear) 
%is important. 
Let us introduce a Kac matrix $\bm$ for a fixed $(h,q)$, 
\[
{\bm}_{\left\{{I'K'M'N'}\right\} \left\{{IKMN}\right\}} 
\equiv \bra{h,q} G^{-}_{N'} G^{+}_{M'} L_{K'} J_{I'} 
  J_{-I} L_{-K} G^{+}_{-M} G^{-}_{-N} \ket{h,q} \,\,.
\]
Each entry of this matrix is a function of $(h,q)$ 
and the central charge $c$. 
%A null state $\ket{\chi}$ which is written as 
%\[
%\ket{\chi} \equiv \sum C_{I_{0} K_{0} M_{0} N_{0}} 
% J_{-{I}_{0}} L_{-{K}_{0}} G^{+}_{-{M}_{0}} G^{-}_{-{N}_{0}} \ket{h,q}
%\]
%is orthogonal to all states $\bra{\xi}$. 
%That is written. 
%\[
%\bra{\xi} = \bra{h,q} G^{-}_{N} G^{+}_{M} L_{K} J_{I} 
%\] 
%Using this state, we obtain the relation.
%\begin{eqnarray*}
%\bracket{\xi}{\chi} &=& 
%  \sum_{\left\{{I_{0} K_{0} M_{0} N_{0}}\right\}} 
%  {C}_{{I}_{0} {K}_{0} {M}_{0} {N}_{0}}
%  \bra{h,q} G^{-}_{N} G^{+}_{M} L_{K} J_{I} 
%  J_{-{I}_{0}} L_{-{K}_{0}} G^{+}_{-{M}_{0}} G^{-}_{-{N}_{0}} \ket{h,q} \\
%&=& \sum_{\left\{{I_{0} K_{0} M_{0} N_{0}}\right\}} 
%  C_{I_{0} K_{0} M_{0} N_{0}} 
% {\bm}_{\left\{{IKMN}\right\} \left\{{I_{0} K_{0} M_{0} N_{0}}\right\}} \\
%&=& 0 \hspace{3cm} \mbox{for all \hspace{3mm}} \left\{{IKMN}\right\}
%\end{eqnarray*}
%This shows that 
The Kac matrix becomes degenerate when at least one null 
state exists. In that case, a determinant of the Kac matrix, 
Kac determinant, vanishes. The  determinant formula of the 
N=2 superconformal field theory is known \cite{BFK}. 
In the Neveu-Shwarz sector,
\begin{eqnarray*}
&&\det {\bm}_{n,m} (\tilde{c} ,h,q) \\
&&= \prod_{\stackrel{\scriptstyle 1 \leq rs \leq 2n}{s \mbox{;} even}}
 {\left({f^{A}_{r,s}}\right)}^{P_{A} (n-\frac{rs}{2} ,m)} 
  \times \prod_{k \in \bhz} {\left({g^{A}_{k}}\right)}^{{
  \tilde{P}}_{A} (n-|k|, m-sgn(k) \mbox{;} k)} \,\,,\\
&&f^{A}_{r,s} (\tilde{c} ,h,q) \equiv 2 (\tilde{c} -1)h- q^{2} -\frac{1}{4} 
  {(\tilde{c} -1)}^{2} +\frac{1}{4} {\left[{(\tilde{c} -1)r +s}\right]}^{2} 
\hspace{1cm}(s \mbox{;even}) \,\,,\\
&&g^{A}_{k}(\tilde{c} ,h,q) \equiv 2h -2kq +(\tilde{c} -1) 
\left({k^{2} -\frac{1}{4}}\right)  
\hspace{2cm}(k \in \bhz) \,\,,\\
&&\tilde{c} \equiv c/3 \,\,,
\end{eqnarray*}
where $n,m$ are level, and relative charge respectively, 
\[
sgn(k)=\left\{
\begin{array}{ll}
+1&(k>0)\\
-1&(k<0) \,\,,
\end{array}
\right.
\]
\begin{eqnarray*}
\sum_{n,m} P_{A} (n,m) {x}^{n} {y}^{m} 
&=& \mathop{\prod}_{k=1}^{\infty} \frac{(1+ {x}^{k-\frac{1}{2}} y)
 (1 +{x}^{k-\frac{1}{2}} {y}^{-1})}{{(1- {x}^{k})}^{2}} \,\,,\\
\sum_{n,m} {\tilde{P}}_{A} (n,m \mbox{;} \hspace{1mm} k) {x}^{n} {y}^{m} 
&=& {(1+ {x}^{|k|} {y}^{sgn(k)})}^{-1} \mathop{\prod}_{k=1}^{\infty} 
 \frac{(1+ {x}^{k-\frac{1}{2}} y)
 (1+ {x}^{k-\frac{1}{2}} {y}^{-1})}{{(1- {x}^{k})}^{2}} \,\,.
\end{eqnarray*}
%This formula provides us with the information where and how many 
%null states appear in the range of $L_{0}$ eigenvalue from $h$ to $h+n$ 
%and of $J_{0}$ eigenvalue from $q$ to $q+m$ beginning with the $\ket{h,q}$ 
%primary field. 

\subsection[Null states]{Factorization of the Kac determinant}

\pr
In representing the N=2 superconformal algebra on the Fock space, the relation 
between the Verma module and the Fock module, and that between the 
dual Verma module and the dual Fock module are very important. 
Particularly whether null states are expressed by the oscillators of 
the free fields is crucial in considering the irreducible representation of 
the N=2 superconformal algebra on the Fock space. In order to study this point, 
let us consider the N=2 superconformal module on the Fock space, 
\begin{eqnarray}
J_{-I} L_{-K} G^{+}_{-M} G^{-}_{-N} \aket 
 =\sum_{I'K'M'N'} R_{\{IKMN\} \{I'K'M'N'\}} (a_{0},{\ma}_{0}) 
 a_{-I'} {\ma}_{-K'} b_{-M'} {\mb}_{-N'} \aket \,\,. \label{eqn:kacr}
\end{eqnarray}
Note that if the determinant of $R_{\{IKMN\} \{I'K'M'N'\} }$ 
vanishes, either there exists some state in the Fock space which 
cannot be expressed by using the $N=2$ generators 
(It means the existence of a cosingular state in $\FOCK$)
or an appropriate linear combination of the state 
(\ref{eqn:kacr}) in the Verma module vanishes
when one rewrites the $N=2$ generators in terms of the  
oscillators (It means the existence of a vanishing null state in $\FOCK$).
The coefficient matrix 
can be expressed as the expectation value of the Fock ground state,
\[
R_{\{IKMN\} \{I'K'M'N'\}}(\alpha ,\bar{\alpha})
 =\abra {\mb}_{N'} b_{M'} {\ma}_{K'} a_{I'} 
  J_{-I} L_{-K} G^{+}_{-M} G^{-}_{-N} \aket \,\,.
\]
%That is to say, 
This gives a mapping from the Verma module to the Fock module, 
\[
\psi \mbox{;} \VERMA \rightarrow \FOCK \,\,.
\]
Similarly, a mapping from the dual Verma module to the dual Fock module \,\,, 
\[
{\psi}^{\ast} \mbox{;} \DVERMA \rightarrow \DFOCK \,\,,
\]
can be considered by defining another coefficient matrix, 
\[
S_{\{I'K'M'N'\} \{IKMN\} }(\alpha ,\bar{\alpha})
 \equiv \abra G^{-}_{N} G^{+}_{M} L_{K} J_{I} 
 a_{-I'} {\ma}_{-K'} b_{-M'} {\mb}_{-N'} \aket \,\,.
\]
If the determinant of $S_{\{I'K'M'N'\} \{IKMN\}}(\alpha ,\bar{\alpha})$ 
vanishes, either an appropriate linear combination of a state 
$\abra G^{-}_{N} G^{+}_{M} L_{K} J_{I}$ vanishes in the dual Fock 
space when one rewrite the $N=2$ generators by the oscillators
(It means the existence of a vanishing null states in $\DFOCK$)
or the equation,
\begin{eqnarray*}
\abra G^{-}_{N} G^{+}_{M} L_{K} J_{I} 
 = \sum_{\{{{I'} {K'} {M'} {N'}}\}} 
	\abra {\mb}_{{N'}} b_{{M'}} {\ma}_{{K'}} a_{{I'}} 
	S_{\{{{I'} {K'} {M'} {N'}}\} 
	 \{{{I} {K} {M} {N}}\}} (\alpha ,\bar{\alpha}) \,\,,
\end{eqnarray*}
cannot be solved inversely and there exists some state in the dual 
Fock space which cannot be expressed by using the $N=2$ generators 
(It means the existence of a cosingular state in $\DFOCK$).
We consider a null state which is non-vanishing in the Fock space, 
\begin{eqnarray*}
\ket{\chi} &=& \sum_{I'K'M'N'} C'_{I'K'M'N'} 
       a_{-I'} {\ma}_{-K'} b_{-M'} {\mb}_{-N'} \aket \\
           & \neq & 0 \,\,.
\end{eqnarray*}
From the null state condition, one obtains a relation,
\begin{eqnarray*}
&&\abra G^{-}_{N} G^{+}_{M} L_{K} J_{I} \ket{\chi} \\
&&=\sum_{I'K'M'N'} C'_{I'K'M'N'} 
   S_{\{I'K'M'N'\} \{IKMN\} }(\alpha ,\bar{\alpha}) \\
&&=0 \,\,.
\end{eqnarray*}
It means that the determinant of $S(\alpha, \bar{\alpha})$ is zero.
Using the same argument, we can show that the determinant of 
$R(\alpha, \bar{\alpha})$ is zero when there exists some null state which is 
non-vanishing in the dual Fock space. 
In addition to these results, we obtain the next important 
relation between these determinants and the Kac determinant. 
The Kac matrix 
${\bm}_{\{I'K'M'N'\} \{IKMN\}}$ is rewritten as, 
\begin{eqnarray*}
\lefteqn{{\bm}_{\{{I_{2} K_{2} M_{2} N_{2}}\} 
 \{{I_{1} K_{1} M_{1} N_{1}}\}}} \\ 
& \equiv & \abra G^{-}_{N_{2}} G^{+}_{M_{2}} L_{K_{2}} J_{I_{2}} 
 J_{-{I}_{1}} L_{-{K}_{1}} G^{+}_{-{M}_{1}} G^{-}_{-{N}_{1}} \aket \\
&=& \sum_{\{{{I'}_{2} {K'}_{2} {M'}_{2} {N'}_{2}}\}} 
  \sum_{\{{{I'}_{1} {K'}_{1} {M'}_{1} {N'}_{1}}\}} 
	\abra {\mb}_{{N'}_{2}} b_{{M'}_{2}} {\ma}_{{K'}_{2}} a_{{I'}_{2}} 
	S_{\{{{I'}_{2} {K'}_{2} {M'}_{2} {N'}_{2}}\} 
	 \{{{I}_{2} {K}_{2} {M}_{2} {N}_{2}}\}} (\alpha ,\bar{\alpha}) \\
& & \mbox{} \times R_{\{{{I}_{1} {K}_{1} {M}_{1} {N}_{1}}\} 
    \{{{I'}_{1} {K'}_{1} {M'}_{1} {N'}_{1}}\}} (\alpha , \bar{\alpha})
     a_{-{I'}_{1}} {\ma}_{-{K'}_{1}} b_{-{M'}_{1}} {\mb}_{-{N'}_{1}} \aket \\
&=& \sum_{\{{IKMN}\}} R_{\{{{I}_{1} {K}_{1} {M}_{1} {N}_{1}}\} \{{IKMN}\}} 
     (\alpha ,\bar{\alpha}) S_{\{{IKMN}\} 
	\{{{I}_{2} {K}_{2} {M}_{2} {N}_{2}}\}} (\alpha ,\bar{\alpha}) \,\,.
\end{eqnarray*}
From this equality, the Kac determinant is rewritten as a product 
of the determinants, 
\[
\det {\bm} =\det R \det S \,\,.
\]
If a null state exists in the Verma module $\VERMA$ or in 
the dual Verma module $\DVERMA$, $\det {\bm} =0$
 and it means that $\det R =0$ or $\det S =0$. \\
Considering these relations, we conclude that \\
\[
\det R =0 \,\,,
\]
when a vanishing null state exists in $\FOCK$ or a non-vanishing null 
state exists in $\DFOCK$. \\
\[
\det S =0 \,\,,
\] 
when a non-vanishing null state exists in $\FOCK$ or a vanishing null state 
exists in $\DFOCK$. \\

A factorization property of the N=2 superconformal Kac determinant has been 
studied before in \cite{Matsu}.
Parametrizing $c,h,q$ by a pair of half integers $(j,k)$ and an integer 
$M \mbox{\hspace{2mm}} (M=2,3,4, \cdots)$ ;
\begin{eqnarray*} 
&&c=3 \left({1-\frac{2}{M}}\right) \,\,,\\
&&h_{j,k} \equiv \frac{1}{M} \left({jk-\frac{1}{4}}\right) \,\,,\\
&&q_{j,k} \equiv \frac{1}{M} (j-k) \mbox{\hspace{2cm}} (j,k \in \bhz) \,\,,
\end{eqnarray*}
the functions $f_{r,s}$ and $g_{s}$  
in the Kac determinant formula can 
be re-expressed as,
\begin{eqnarray}
&&f_{r,s} =\frac{1}{M^{2}} \left({r-\frac{M}{2} s +j+k}\right)
  \left({r- \frac{M}{2} s -j-k}\right) \hspace{0.8cm}
  (1 \leq rs \leq 2n,\mbox{\hspace{2mm}} s \mbox{;even}) \,\,, \label{eqn:ff}\\
&&g_{s} =-\frac{2}{M} (s+j)(s-k) \hspace{5.5cm} (s \in \bhz) \,\,.\label{eqn:gg}
\end{eqnarray}
In the formula (\ref{eqn:ff}), null states appear at level $rs/2$ and 
relative charge $0$. Taking a primary state with a pair of quantum numbers 
$(j,k)$, null states corresponding to the 1st parenthesis of (\ref{eqn:ff}) 
have quantum numbers, 
\begin{eqnarray}
(M \tilde{s} -k, M \tilde{s} -j) \hspace{2cm} 
(\tilde{s} \equiv s/2) \,\,, \label{eqn:n1}
\end{eqnarray} 
because of the relation $f_{j,k} +r \tilde{s} =
 \frac{1}{M} \left[{(M \tilde{s} -k)(M \tilde{s} -j) -\frac{1}{4}}\right]$. 
These null states appear at the level $\tilde{s} (M \tilde{s} -j-k)$ and
relative charge $0$. The null states
appearing at the level $1$ and relative charge $0$ corresponds to the 
condition 
$\tilde{s} =1, \mbox{\hspace{1.5mm}} M=2,\mbox{\hspace{1.5mm}} j=k=1/2$ 
and we verified that this state is non-vanishing in the Fock space by 
using the representation (\ref{eqn:A3}) in the Appendix.
We also verified that the null state appearing at level $2$ 
and relative charge $0$ associated with the condition
$\tilde{s}=1, \mbox{\hspace{1.5mm}} M-j-k=2$ is non-vanishing in the Fock 
space by explicit calculations.
Therefore we suggest that null states $v_{2 \tilde{s} -1}$ with the 
quantum number (\ref{eqn:n1}) are associated to the zero of the determinant of 
$S$.
Similarly null states corresponding to the 2nd parenthesis of 
(\ref{eqn:ff}) have quantum numbers,
\begin{eqnarray}
(j+M \tilde{s} ,k+M \tilde{s}) \mbox{\hspace{1cm}} 
(\tilde{s} \equiv s/2) \,\,, \label{eqn:n2}
\end{eqnarray}
because of the relation 
$h_{j,k} +r \tilde{s} 
 =\frac{1}{M} \left[{(j+M \tilde{s})(k+M \tilde{s}) -\frac{1}{4}}\right]$. 
The null states associated with quantum numbers (\ref{eqn:n1}),(\ref{eqn:n2})
appear one after the other. Thus we suggest that the cosingular state 
$v_{2 \tilde{s}}$ associated with (\ref{eqn:n2}) corresponds to the zero of the 
determinant of $R$. Taking these consideration into account, we suggest
that the arrows of the embedding diagram in the Fock space flow out of
the states $v_{2 \tilde{s}}$ and flow into the states $v_{2 \tilde{s} -1}$
or $v_{2 \tilde{s} +1}$.
On the other hand, in the formula (\ref{eqn:gg}), one null state 
corresponding to (\ref{eqn:gg}) appears at relative charge $+1$ and level $k$,
\begin{eqnarray}
h_{j,k} +k =\frac{1}{M} \left[{(j+M)k -\frac{1}{4}}\right] \,\,.\label{eqn:n3}
\end{eqnarray}
From this relation, the quantum number of this 
null state become $(j+M,k)$. 
We constructed null states corresponding to 
the condition (\ref{eqn:n3}) explicitly 
( eg. (\ref{eqn:A1})(\ref{eqn:A4}) in the Appendix )
and verified that these null states vanish in the Fock space. 
Thus some cosingular state $w_{1}$ associated with (\ref{eqn:n3}) exists and 
the arrow in the embedding diagram in the Fock space flows out of $w_{1}$
into $v_{0}$.
We note that other null states 
appear at relative charge $-1$ 
and level $j$, 
\begin{eqnarray}
h_{j,k} +j =\frac{1}{M} \left[{j(k+M) -\frac{1}{4}}\right] \label{eqn:n4} \,\,.
\end{eqnarray}
From this relation, we see that 
the quantum number of this state is $(j,k+M)$. 
By the explicit calculation 
( eq. (\ref{eqn:A2})(\ref{eqn:A5}) in the Appendix ), 
some cosingular state $u_{1}$ associated with (\ref{eqn:n4}) exists and the 
arrow in the embedding diagram in the Fock space flows out of $u_{1}$
into $v_{0}$. 
Furthermore one can construct (co)singular states from $w_{1}$ and $u_{1}$.
We denote the states constructed from $w_{1}$ as $w_{2 \tilde{s}}$ and
$w_{2 \tilde{s} +1}$. 
The states $w_{2 \tilde{s}}$ has a set of quantum numbers,
\[
(M(\tilde{s} +1) -k, M \tilde{s} -j) \,\,,
\]
and is obtained by using the relation in 
the 1st parenthesis of (\ref{eqn:ff}).
On the other hand, the 
state $w_{2 \tilde{s} +1}$ has a set of quantum numbers,
\[
(j+M \tilde{s} , k+M \tilde{s}) \,\,,
\] 
and is obtained by using the relation in 
the 2nd parenthesis of (\ref{eqn:ff}).
Assuming that the number of arrows with opposite directions 
compared with the directions of arrows in the Verma embedding diagram is 
independent of the choice of paths which one follows from one state to 
the other state, we obtain an embedding diagram of the Fock space of an 
$N=2$ superconformal model. \\
Conjecture : Embedding diagram of the Fock space \hspace{2mm}
($j>0$ \hspace{1mm} and \hspace{1mm} $k>0$)
\[
\begin{array}{ccccccccc}
\cdots & \yh & u_{3} & \ym & u_{2} & \yh & u_{1} &      &       \\
       & \yd &       & \yd &       & \yd &       & \msi &       \\
\cdots & \ym & v_{3} & \yh & v_{2} & \ym & v_{1} & \yh  & v_{0} \\
       & \yu &       & \yu &       & \yu &       & \mue &       \\
\cdots & \yh & w_{3} & \ym & w_{2} & \yh & w_{1} &      & 
\end{array}
\]
%\vspace{5cm}
\section{BRST analysis}

\pr
In this section, the BRST operators are considered. 
These BRST operators provide us with a recipe to define an 
irreducible representation space of the N=2 superconformal algebra on the 
Fock space as a cohomology class of 
these BRST operators. 
At the end of this section, we obtain a character formula 
as an application of the BRST analysis.
\subsection[BRST analysis]{Actions of the BRST operators on the Fock spaces}

\pr
Let us recall that we parametrized the eigenvalues of $L_{0}$ and 
$J_{0}$, $(h,q)$ by using half integers $(j,k)$,
\begin{eqnarray*}
&&h_{j,k} \equiv \frac{1}{M} \left({jk-\frac{1}{4}}\right) \,\,, \\
&&q_{j,k} \equiv \frac{1}{M} \left({j-k}\right) \,\,,\\
&&c \equiv 3 \left({1-\frac{2}{M}}\right)\,\,.
\end{eqnarray*} 
As is well known, in order to obtain the irreducible representation, 
one has to restrict the range of $(j,k)$ to 
the following regions \cite{BFK,Matsu},  
\begin{eqnarray*}
&&0 < j < M \,\,,\\
&&0 < k < M \,\,,\\
&&0 < j+k < M \mbox{\hspace{1cm}} (j,k \in \bhz)\,\,.
\end{eqnarray*} 
This restriction is owing to the existence of null states which one must 
subtract to get a positive-definite theory. 
A BRST analysis provides us with a method of subtracting null states 
and explains the necessity for restricting the range of $(j,k)$. 

We first recall that  the $Q_{f}$ operator maps a 
Fock space ${\FOCK}_{j,k}$ to ${\FOCK}_{j,k+M}$. 
Secondly by the action of the ${\bar{Q}}_{f}$ operator, a Fock space 
${\FOCK}_{j,k}$ is mapped to ${\FOCK}_{j+M,k}$.
On the other hand, the bosonic BRST operator $Q^{(n)}_{B}$ maps 
${\FOCK}_{j,k}$ to a Fock space 
${\FOCK}_{j-n,k-n}$. 
These BRST operators have a value $(h,q)=(0,0)$. Moreover 
they (anti)commute with all generators of the N=2 superconformal algebra. 
With these properties, the (co)singular states in ${\FOCK}_{j,k}$ 
are mapped to (co)singular states in 
${\FOCK}_{j,k+M}, {\FOCK}_{j+M,k}$, or ${\FOCK}_{j-n,k-n}$ 
by actions of $Q_{f} ,{\bar{Q}}_{f}$, or $Q^{(n)}_{B}$, 
respectively, if their images are non-vanishing. Thus null states could be 
constructed by using these BRST operators. 
From the start, we restrict the region of $(j,k)$ 
to the region with $jk > 0$ by 
hand. In fact for $jk <0$, the eigenvalue of $L_{0}$, $h_{j,k}$ is negative 
and states in the region with $jk <0$ don't satisfy the unitarity relation. 

To begin with, let us recall the embedding diagram 
we have conjectured in section 4.2.\\
An embedding diagram in the Fock space \hspace{2mm}
($j>0$ \hspace{1mm} and \hspace{1mm} $k>0$) is given by
\[
\begin{array}{ccccccccl}
\cdots & \yh & u_{3} & \ym & u_{2} & \yh & u_{1} &      &       \\
       & \yd &       & \yd &       & \yd &       & \msi &       \\
\cdots & \ym & v_{3} & \yh & v_{2} & \ym & v_{1} & \yh  & v_{0} \\
       & \yu &       & \yu &       & \yu &       & \mue &       \\
\cdots & \yh & w_{3} & \ym & w_{2} & \yh & w_{1} &      & \mbox{\hspace{30mm}}.
\end{array}
\]
In this diagram, an arrow, or a chain of arrows, goes from one vector 
to another whenever the latter is in the submodule generated by the former. 
Especially a state $v_{i+1} \mbox{\hspace{3mm}} (i=1,2,3,\cdots)$ 
can be constructed 
from the state $u_{i} \mbox{\hspace{3mm}} (i=1,2,3, \cdots)$ together 
with the state $w_{i} \mbox{\hspace{3mm}} (i=1,2,3, \cdots)$ and cannot be 
constructed from the only one of the two states $u_{i} , w_{i}$. 
Each state is parametrized by a pair of numbers $(h,q)$ or 
a pair of half integers $(j,k)$. 
If $v_{0}$ is in a Fock space 
${\FOCK}_{j,k}  \mbox{\hspace{2mm}} (j>0 $ and $k>0)$, 
%$(i.e. \mbox{\hspace{2mm}} 
%(h,q)= \left({\frac{1}{M} \left({jk-\frac{1}{4}}\right), 
%\mbox{\hspace{2mm}} \frac{1}{M} (j-k)}\right)$, 
the following quantum numbers are 
assigned to the null states, 
\[
\begin{array}{lclcl}
v_{2i}   & : &(j+Mi, k+Mi)     & \hspace{1cm} &(i=0,1,2, \cdots) \,\,,\\
v_{2i-1} & : &(Mi-k, Mi-j)     & \hspace{1cm} &(i=1,2,3, \cdots) \,\,,\\
w_{2i+1} & : &(j+M(i+1), k+Mi) & \hspace{1cm} &(i=0,1,2, \cdots) \,\,,\\
w_{2i}   & : &(M(i+1)-k, Mi-j) & \hspace{1cm} &(i=1,2,3, \cdots) \,\,,\\
u_{2i+1} & : &(j+Mi, k+M(i+1)) & \hspace{1cm} &(i=0,1,2, \cdots) \,\,,\\
u_{2i}   & : &(Mi-k, M(i+1)-j) & \hspace{1cm} &(i=1,2,3, \cdots) \,\,.
\end{array}
\]

On the other hand, an embedding diagram of the Fock space in the case of 
$j<0$ and $k<0$ is drawn by reversing all the arrows in the diagram of 
the above case. \\
An embedding diagram in the Fock space \hspace{2mm}
($j<0$ \hspace{1mm} and \hspace{1mm} $k<0$) reads,
\[
\begin{array}{ccccccccl}
\cdots & \ym & u_{3} & \yh & u_{2} & \ym & u_{1} &      &       \\
       & \yu &       & \yu &       & \yu &       & \hue &       \\
\cdots & \yh & v_{3} & \ym & v_{2} & \yh & v_{1} & \ym  & v_{0} \\
       & \yd &       & \yd &       & \yd &       & \hsi &       \\
\cdots & \ym & w_{3} & \yh & w_{2} & \ym & w_{1} &      & \mbox{\hspace{30mm}}. 
\end{array}
\]

In the case of the $(j>0 \mbox{\hspace{2mm}} k>0)$ diagram, 
let us investigate actions of the BRST operators 
on the Fock space. \\
 Case 1  Fermionic BRST operator ${\bar{Q}}_{f}$ \\
This operator maps a state in the Fock space ${\FOCK}_{j,k}$ to a state 
in the Fock space ${\FOCK}_{j+M,k}$. 
Frist, let us consider ${\bar{Q}}_{f} v^{(j,k)}_{0}$ where 
$v^{(j,k)}_{0}$ is a primary state in the Fock space ${\FOCK}_{j,k}$. 
Naively ${\bar{Q}}_{f} v^{(j,k)}_{0}$ is a state with 
$(h,q)=\left({\frac{1}{M} \left({jk-\frac{1}{4}}\right) 
, \mbox{\hspace{2mm}} \frac{1}{M} (j-k)}\right)$ because the BRST operator 
${\bar{Q}}_{f}$ has $(h,q)=(0,0)$, but in fact there is no such a state with 
the same $(h,q)$ as ${\bar{Q}}_{f} v^{(j,k)}_{0}$ in the Fock space 
${\FOCK}_{j+M,k}$. From this reason, ${\bar{Q}}_{f} v^{(j,k)}_{0}$  
vanishes. Similarly ${\bar{Q}}_{f} v^{(j,k)}_{1}$ is vanishing. 
Secondly the states $u^{(j,k)}_{i} \mbox{\hspace{3mm}} (i=1,2,3, \cdots)$ 
in the Fock space ${\FOCK}_{j,k}$ all have the same $U(1)$ charge 
$q=\frac{1}{M} (j-k) -1$, and the states ${\bar{Q}}_{f} u^{(j,k)}_{i}$ 
should have $U(1)$ charge $q=\frac{1}{M} (j-k) -1$. 
However, there is no such a state with this value of $U(1)$ 
charge in the Fock space 
${\FOCK}_{j+M,k}$. Thus ${\bar{Q}}_{f} u^{(j,k)}_{i}$ also vanishes. 
As for $w^{(j,k)}_{1}$ in the Fock space ${\FOCK}_{j,k}$,  
${\bar{Q}}_{f} w^{(j,k)}_{1}$ is non-vanishing and should be a primary state 
$v^{(j+M,k)}_{0}$ in the Fock space ${\FOCK}_{j+M,k}$ . 
Here we introduce a symbol for convenience, 
\begin{eqnarray*}
\lefteqn{{\ld}^{(R)}_{-(N)} \equiv } \\
&& \{ J_{-{m}_{1}} \cdots J_{-{m}_{r}} 
  L_{-{n}_{1}} \cdots L_{-{n}_{s}} 
  G^{+}_{-{k}_{1}} \cdots G^{+}_{-{k}_{t}} 
  G^{-}_{-{l}_{1}} \cdots G^{-}_{-{l}_{u}} 
  \mbox{; \hspace{2mm}} m,n,k,l >0 \,\,,\\
&& \hspace*{1cm}   with  \hspace{3mm} t-u= R  
 \hspace{1cm}   ( i.e. \mbox{relative charge is } R ) \,\,,\\
&& \hspace*{1.2cm}   and \hspace{2mm}  
   N=\sum_{i=1}^{r} {m}_{i} +\sum_{i=1}^{s} {n}_{i} 
   +\sum_{i=1}^{t} {k}_{i} +\sum_{i=1}^{u} {l}_{i} 
   \hspace{1cm}   ( i.e.\mbox{ level is } N) \hspace{3mm} \} \,\,.
\end{eqnarray*}
From the embedding diagram, one constructs the state $w^{(j,k)}_{2}$ 
from the state $w^{(j,k)}_{1}$ by using some element 
$f^{(0)}_{1} \in {\ld}^{(0)}_{-(2(M-k-j))}$ ;
\[
w^{(j,k)}_{2} =f^{(0)}_{1} w^{(j,k)}_{1} \,\,.
\] 
Acting the BRST operator ${\bar{Q}}_{f}$ on this state $w^{(j,k)}_{2}$, 
we get the relation,
\begin{eqnarray*}
{\bar{Q}}_{f} {w}^{(j,k)}_{2} &=& {\bar{Q}}_{f} f^{(0)}_{1} w^{(j,k)}_{1} \\
 &=& f^{(0)}_{1} {\bar{Q}}_{f} w^{(j,k)}_{1} \\
 &=& f^{(0)}_{1} v^{(j+M,k)}_{0} \\
 &=& v^{(j+M,k)}_{1} \,\,.
\end{eqnarray*} 
The last equality is obtained because the state 
$f^{(0)}_{1} v^{(j+M,k)}_{0}$ is a null state in the Fock space 
${\FOCK}_{j+M,k}$ with relative charge $0$ and level $2(M-k-j)$. 
Similarly the state $v^{(j,k)}_{2}$ is constructed from 
$w^{(j,k)}_{1}$ and $u^{(j,k)}_{1}$ by using appropriate operators 
$f^{(-1)}_{1} \in {\ld}^{(-1)}_{-(M+j)}$ and 
${f'}^{(+1)}_{1} \in {\ld}^{(+1)}_{-(M+k)}$ \,\,, 
\[
v^{(j,k)}_{2} = f^{(-1)}_{1} w^{(j,k)}_{1} +{f'}^{(+1)}_{1} u^{(j,k)}_{1} \,\,.
\]
Acting the BRST operator ${\bar{Q}}_{f}$ on this state, one obtains 
the relation,
\begin{eqnarray*}
{\bar{Q}}_{f} {v}^{(j,k)}_{2}
&=& - {f}^{(-1)}_{1} {\bar{Q}}_{f} w^{(j,k)}_{1} 
    - {f'}^{(+1)}_{1} {\bar{Q}}_{f} u^{(j,k)}_{1} \\
&=& - {f}^{(-1)}_{1} {\bar{Q}}_{f} w^{(j,k)}_{1} \\
&=& - {f}^{(-1)}_{1} v^{(j+M,k)}_{0} \,\,.
\end{eqnarray*}
The state $f^{(-1)}_{1} v^{(j+M,k)}_{0}$ appearing in 
the last equality should be a state with the relative charge $-1$ and 
the level $M+j$ and a descendant of the primary field $v^{(j+M,k)}_{0}$ . 
But such a descendant with this condition in the Fock space 
${\FOCK}_{j+M,k}$ doesn't exist 
( the state $u^{(j+M,k)}_{1}$ in the Fock space ${\FOCK}_{j+M,k}$ has 
the same quantum number, but this state is not a descendant of 
a primary state as is seen from the direction of the arrow from $u_{1}$ 
to $v_{0}$ ). 
From these consideration, one understands that 
${\bar{Q}}_{f} v^{(j,k)}_{2}$ vanishes. 
With the same considerations, we obtain actions of the BRST operator 
${\bar{Q}}_{f}$ on the states in ${\FOCK}_{j,k}$, 
\begin{eqnarray*}
&&{\bar{Q}}_{f} v^{(j,k)}_{i} =0 \hspace{2cm} (i=0,1,2, \cdots) \,\,,\\ 
&&{\bar{Q}}_{f} u^{(j,k)}_{i} =0 \hspace{2cm} (i=1,2,3, \cdots) \,\,,\\
&&{\bar{Q}}_{f} w^{(j,k)}_{i} = v^{(j+M,k)}_{i-1} \hspace{1cm} 
 (i=1,2,3, \cdots) \,\,.
\end{eqnarray*}
\\
Case 2 Fermionic BRST operator $Q_{f}$ \\
In this case, the arguments proceed in the same way as the Case 1 
with following replacements,
\begin{eqnarray*}
{\bar{Q}}_{f}   & \rightarrow & Q_{f} \,\,,\\
{\FOCK}_{j+M,k} & \rightarrow & {\FOCK}_{j,k+M} \,\,,\\
{\ld}^{(\pm 1)}_{-(Mi+k)} & \rightarrow & {\ld}^{(\pm 1)}_{-(Mi+j)} \,\,,\\
u^{(j,k)}_{i} & \rightarrow &  w^{(j,k)}_{i} \,\,,\\
w^{(j,k)}_{i} & \rightarrow &  u^{(j,k)}_{i} \,\,,\\
v^{(j+M,k)}_{0} & \rightarrow &  v^{(j,k+M)}_{0} \,\,.
\end{eqnarray*}
We obtain actions of the BRST operator $Q_{f}$ on the states in 
${\FOCK}_{j,k}$,
\begin{eqnarray*}
&&Q_{f} v^{(j,k)}_{i} =0 \hspace{2cm} (i=0,1,2, \cdots) \,\,,\\
&&Q_{f} w^{(j,k)}_{i} =0 \hspace{2cm} (i=1,2,3, \cdots) \,\,,\\
&&Q_{f} u^{(j,k)}_{i} =v^{(j,k+M)}_{i-1} \hspace{1cm} (i=1,2,3, \cdots) \,\,. 
\end{eqnarray*}
\\
Case 3 Bosonic BRST operator $Q^{(n)}_{B}$ \\
We recall the property of $Q^{(n)}_{B}$, 
\begin{eqnarray*}
&&Q^{(n=M)}_{B} =0 \,\,,\\
&&0< n=j+k+Ml <M \hspace{1cm} (\mbox{for some \hspace{2mm}}  l \in \bz) \,\,,
\end{eqnarray*}
when acting on ${\FOCK}_{j,k}$.
We fix the range of $(j,k)$ in view of the above property of $Q_{B}$,
\begin{eqnarray*}
&&0 < j < M \,\,,\\
&&0 < k < M \,\,,\\
&&0 < j+k < M \,\,.
\end{eqnarray*}
Then a semi-infinite sequence of complex (including a Fock space 
${\FOCK}_{j,k}$ at which the sequence terminates ) should exist, 
\begin{eqnarray*}
&&{\FOCK}_{j,k} \ya{M-j-k} {\FOCK}_{M-k,M-j} \ya{j+k} 
 {\FOCK}_{j+M,k+M} \ya{M-j-k} \cdots \\
&&\cdots \ya{j+k} {\FOCK}_{j+Ml,k+Ml} \ya{M-j-k} 
 {\FOCK}_{M(l+1)-k,M(l+1)-j} \ya{j+k} \cdots \mbox{\hspace{20mm}}. 
\end{eqnarray*}
In order to consider the action of the BRST operator $Q_{B}$ on the states, 
we take the Fock space ${\FOCK}_{j+Ml,k+Ml} \hspace{2mm} (l=1,2,3, \cdots)$.  
(Arguments are the same in the other cases.) The operator $Q^{(j+k)}_{B}$ 
maps the Fock space ${\FOCK}_{j+Ml,k+Ml}$ to  
${\FOCK}_{Ml-k,Ml-j}$. 
Firstly one shows that,
\begin{eqnarray*}
&&Q^{(j+k)}_{B} v^{(j+Ml,k+Ml)}_{0} = v^{(Ml-k,Ml-j)}_{1} \,\,,\\
&&Q^{(j+k)}_{B} w^{(j+Ml,k+Ml)}_{1} = w^{(Ml-k,Ml-j)}_{2} \,\,,\\
&&Q^{(j+k)}_{B} u^{(j+Ml,k+Ml)}_{1} = u^{(Ml-k,Ml-j)}_{2} \,\,.
\end{eqnarray*}  
From the embedding diagram, the state 
$v^{(j+Ml,k+Ml)}_{1} \in {\FOCK}_{j+Ml,k+Ml}$ is constructed from 
the primary field in the ${\FOCK}_{j+Ml,k+Ml}$ by using an appropriate 
operator $f^{(0)}_{0} \in {\ld}^{(0)}_{-((2l+1)(M-j-k))} $, 
\[
v^{(j+Ml,k+Ml)}_{1}= f^{(0)}_{0} v^{(j+Ml,k+Ml)}_{0} \,\,.
\]
We get the action of $Q^{(j+k)}_{B}$ on this state,
\begin{eqnarray*}
Q^{(j+k)}_{B} v^{(j+Ml,k+Ml)}_{1} 
&=& f^{(0)}_{0} Q^{(j+k)}_{B} v^{(j+Ml,k+Ml)}_{0} \\
&=& f^{(0)}_{0} v^{(Ml-k,Ml-j)}_{1} \,\,.
\end{eqnarray*}
The state $f^{(0)}_{0} v^{(Ml-k,Ml-j)}_{1}$ appearing in the last equality 
should be a state with the $U(1)$ charge $\frac{1}{M} (j-k)$ and conformal 
weight 
$\frac{1}{M} \left[{ \{ {M(l+1)-k} \} \{{M(l+1)-j}\} -\frac{1}{4}} \right]$
 and furthermore a descendant of the state $v^{(Ml-k,Ml-j)}_{1}$ in the 
Fock space ${\FOCK}_{Ml-k,Ml-j}$ .
But such a descendant with the same $(h,q)$ as this state doesn't exist. 
( The state $v^{(Ml-k,Ml-j)}_{2}$ has the same $(h,q)$ but this is not a 
descendant of the state $v^{(Ml-k,Ml-j)}_{1}$ as is seen from the direction 
of the arrow from $v_{2}$ to $v_{1}$. ) 
From this consideration, $Q^{(j+k)}_{B} v_{1}$ vanishes. 
Thirdly the state $w^{(j+Ml,k+Ml)}_{2}$ is expressed by 
using some operator, 
$f^{(0)}_{1} \in {\ld}^{(0)}_{-((2l+2)(M-j-k))} $,  
\[
w^{(j+Ml,k+Ml)}_{2} =f^{(0)}_{1} w^{(j+Ml,k+Ml)}_{1} \,\,,
\]
 and we obtain the action of the $Q^{(j+k)}_{B}$ on this, 
\begin{eqnarray*}
& & Q^{(j+k)}_{B} w^{(j+Ml,k+Ml)}_{2} \\
&=& f^{(0)}_{1} Q^{(j+k)}_{B} w^{(j+Ml,k+Ml)}_{1} \\
&=& f^{(0)}_{1} w^{(Ml-k,Ml-j)}_{2} \,\,.
\end{eqnarray*}
But such a descendant doesn't exist in the Fock module 
${\FOCK}_{Ml-k,Ml-j}$.  
( The state $w^{(Ml-k,Ml-j)}_{3}$ has the same $(h,q)$ but is not 
a descendant of $w^{(Ml-k,Ml-j)}_{2}$ and is excluded in the candidate. )
From this consideration, we obtain the relation, 
\[
Q^{(j+k)}_{B} w^{(j+Ml,k+Ml)}_{2} =0 \,\,.
\] 
A similar consideration is applied to the state $u_{2}$ and 
we obtain,
\[
Q^{(j+k)}_{B} u^{(j+Ml,k+Ml)}_{2} =0 \,\,.
\] 
Repeating the same arguments, we obtain the following results, 
\begin{equation}
\begin{array}{lclr}
Q^{(j+k)}_{B} v^{(j+Ml,k+Ml)}_{2i} & = & v^{(Ml-k,Ml-j)}_{2i+1} & 
 (i=0,1,2, \cdots) \,\,,\\
Q^{(j+k)}_{B} w^{(j+Ml,k+Ml)}_{2i-1} & = & w^{(Ml-k,Ml-j)}_{2i} &
 (i=1,2,3, \cdots) \,\,,\\
Q^{(j+k)}_{B} u^{(j+Ml,k+Ml)}_{2i-1} & = & u^{(Ml-k,Ml-j)}_{2i} &
 (i=1,2,3, \cdots) \,\,,\\
Q^{(j+k)}_{B} v^{(j+Ml,k+Ml)}_{2i+1} & = & 0 &
 (i=0,1,2, \cdots) \,\,,\\
Q^{(j+k)}_{B} w^{(j+Ml,k+Ml)}_{2i} & = & 0 & 
 (i=1,2,3, \cdots) \,\,,\\
Q^{(j+k)}_{B} u^{(j+Ml,k+Ml)}_{2i} & = & 0 &
 (i=1,2,3, \cdots) \,\,.\\
\end{array}
\end{equation}
By the same arguments applied to the 
${\FOCK}_{M(l+1)-k,M(l+1)-j}$, the states 
$v^{(j+Ml,k+Ml)}_{2i} ,w^{(j+Ml,k+Ml)}_{2i}$ , and $u^{(j+Ml,k+Ml)}_{2i}$ 
can be expressed as BRST ( $Q^{(M-j-k)}_{B}$ ) exact states, 
\begin{eqnarray*}
v^{(j+Ml,K+Ml)}_{2i+1} &=& Q^{(M-j-k)}_{B} v^{(M(l+1)-k,M(l+1)-j)}_{2i} \,\,,\\
w^{(j+Ml,k+Ml)}_{2i} &=& Q^{(M-j-k)}_{B} w^{(M(l+1)-k,M(l+1)-j)}_{2i-1} \,\,,\\
u^{(j+Ml,k+Ml)}_{2i} &=& Q^{(M-j-k)}_{B} u^{(M(l+1)-k,M(l+1)-j)}_{2i-1} \,\,.
\end{eqnarray*}
These relations are consistent with the nilpotency of 
the BRST operator $Q_{B}$,
\[
Q^{(j+k)}_{B} Q^{(M-j-k)}_{B} = 0 \,\,.
\]
Figures 1, 2 and 3 illustrate these relations. 
\subsection[BRST analysis]{Construction of the irreducible representation} 

\pr
In this subsection, we define the irreducible representation space of an N=2 
superconformal module on the Fock space as BRST cohomology classes. 
Then we rederive the character formula of the N=2 superconformal minimal 
unitary models \cite{Matsu,Dob}. 

To begin with, let us consider what we should do to obtain the 
irreducible representation space of the N=2 superconformal module on the Fock 
space. In Verma module, one must subtract the null states to obtain the 
irreducible representation space. \\ 
%\vspace{5cm}
An embedding diagram in the Verma module is given by 
\[
\begin{array}{ccccccccl}
\cdots & \yh & \sh{3} & \yh & \sh{2} & \yh & \sh{1} &      &       \\
       & \yl &        & \yl &        & \yl &        & \hue &       \\
\cdots & \yh & s_{3}  & \yh & s_{2}  & \yh & s_{1}  & \yh  & s_{0} \\
       & \yl &        & \yl &        & \yl &        & \hsi &       \\
\cdots & \yh & \st{3} & \yh & \st{2} & \yh & \st{1} & & \mbox{\hspace{30mm}}. 
\end{array}
\]
\\
In the above embedding diagram of the Verma module, we have only to 
subtract three null states $s_{1} ,s^{+}_{1}$, and $s^{-}_{1}$ 
because the other null states are descendants of these null states. 
Let us consider the mapping which maps the states in the Verma module 
to the states in the Fock module (We re-express 
the states in the Verma module using the oscillators),
\[
\psi \mbox{;} {\VERMA} \rightarrow {\FOCK} \,\,.
\] 
Comparing the directions of arrows in the Verma module embedding diagram 
with those in the Fock module embedding diagram, we obtain the result,
\begin{eqnarray*}
&&\psi ( s_{0} ) = v_{0} \,\,,\\
&&\psi ( s_{1} ) = v_{1} \,\,,\\
&&\psi ( s^{+}_{1} ) =0 \,\,,\\
&&\psi ( s^{-}_{1}) =0 \,\,.
\end{eqnarray*}

On the other hand, we must also consider the dual Verma module. \\ 
%\vspace{5cm}
An embedding diagram in the dual Verma module reads,
\[
\begin{array}{ccccccccl}
\cdots & \ym & \sha{3} & \ym & \sha{2} & \ym & \sha{1} &      &        \\
       & \yr &         & \yr &         & \yr &         & \msi &        \\
\cdots & \ym & \sa{3}  & \ym & \sa{2}  & \ym & \sa{1}  & \ym  & \sa{0} \\
       & \yr &         & \yr &         & \yr &         & \mue &        \\
\cdots & \ym & \sta{3} & \ym & \sta{2} & \ym & \sta{1} & &\mbox{\hspace{30mm}}. 
\end{array}
\]
\\
In the dual Verma module, we have only to subtract 
the contribution of the three states 
${s}^{\ast}_{1}$, $({s}^{+}_{1})^{\ast}$, and $({s}^{-}_{1})^{\ast}$ . 
One also considers a mapping from 
the Fock module to the dual Verma module 
(One re-expresses the states written by the oscillators 
using the generators of the N=2 superconformal algebra),
\[
{\psi}^{\ast} \mbox{;} {\FOCK} \rightarrow {\DVERMA} \,\,.
\] 
Comparing the directions of arrows in the Fock space embedding 
diagram to those in the dual Verma module embedding diagram, 
we obtain the relation,
\begin{eqnarray*}
&&{\psi}^{\ast} ({v_{0}}) =s^{\ast}_{0} \,\,,\\
&&{\psi}^{\ast} ({v_{1}}) =0 \,\,,\\
&&{\psi}^{\ast} ({w_{1}}) =({s}^{+}_{1})^{\ast} \,\,,\\
&&{\psi}^{\ast} ({u_{1}}) = ({s}^{-}_{1})^{\ast} \,\,.
\end{eqnarray*}
Together with the study of $\psi$, one notices that one has only to subtract 
the contributions of $v_{1} ,w_{1} ,u_{1}$ in the Fock module to obtain the 
irreducible representation spaces. 
In order to subtract these states, we use the BRST operators. 

Firstly we subtract the contributions of $w_{1}$ and $u_{1}$ by using 
the fermionic BRST operator $Q_{F}$,
\[
Q_{F} \equiv Q_{f} + {\bar{Q}}_{f} \,\,.
\]
Nilpotency of this operator is evident from the fact, 
\[
Q^{2}_{f} = {\bar{Q}}^{2}_{f} = \ac{Q_{f}}{{\bar{Q}}_{f}} =0 \,\,.
\]
Also this operator satisfys the relation,
\[
 \cm{{Q}_{F}}{{Q}_{B}} =0 \,\,.
\]
Next we define the complexes $\ft{(\ast)}{+} ,\ft{(\ast)}{-}$ as follows, 
\begin{eqnarray*}
&&\ft{(n)}{+} \equiv 
 {\FOCK}_{j+Mn,k} \bigoplus {\FOCK}_{j+M(n-1),k+M} \bigoplus 
 {\FOCK}_{j+M(n-2),k+2M} \bigoplus \cdots \bigoplus {\FOCK}_{j,k+Mn} \,\,,\\
&&\ft{(n)}{-} \equiv 
 {\FOCK}_{M(n+1)-k,M-j} \bigoplus {\FOCK}_{Mn-k,2M-j} \bigoplus 
 {\FOCK}_{M(n-1)-k,3M-j} \bigoplus \cdots \bigoplus 
     {\FOCK}_{M-k,M(n+1)-j}\,\,, \\ 
&& \hspace*{3cm}(n=0,1,2, \cdots) \,\,.
\end{eqnarray*}
Then the BRST operator $Q_{F}$ acts on $\ft{(n)}{\pm}$,
\[
Q^{\left[{n}\right]}_{F} \mbox{;} 
 {\ft{(n)}{\pm}} \rightarrow {\ft{(n+1)}{\pm}} \,\,.
\] 
According to the results obtained in the previous subsection, the following 
relations are satisfied for both $\ft{(n)}{+}$ and $\ft{(n)}{-}$ \,\,,
\[
\ker Q^{\left[{n}\right]}_{F} =\left\{
\begin{array}{l}
\vv{j+Mn,k} \bigoplus \vv{j+M(n-1),k+M} \bigoplus 
 \cdots \bigoplus \vv{j,k+Mn} \\
\vv{M(n+1)-k,M-j} \bigoplus \vv{Mn-k,2M-j} \bigoplus 
 \cdots \bigoplus \vv{M-k,M(n+1)-j} 
\end{array}
\right. \,\,,
\]
\[
\vv{l,m} \equiv 
\left\{ {\mbox{state in} \hspace{2mm} {\FOCK}_{l,m} \hspace{2mm} 
  \mbox{generated by} 
 \{ {v_{0} ,v_{1} ,v_{2} , \cdots} \} } \right\}  \,\,.
\]
In addition, We obtain the relation,
\[
\ker Q^{\left[{n}\right]}_{F} = \mbox{Im}  
  Q^{\left[{n-1}\right]}_{F} \hspace{5mm} 
 \mbox{for \hspace{2mm}} n \geq 1 \,\,.
\] 
In short, the cohomology classes ${\pmh}^{(n)}_{Q_{F}}$ are as follows, 
\[
{\th}^{(n)}_{Q_{F}} = \left\{
\begin{array}{ll}
\vv{j,k} & (n=0) \\
 0       & (n \neq 0) 
\end{array}
\right. \,\,,
\]
\[
{\hh}^{(n)}_{Q_{F}} = \left\{
\begin{array}{ll}
\vv{M-k,M-j} & (n=0) \\
 0           & (n \neq 0)
\end{array}
\right. \,\,.
\]
Thus we can subtract the states $w_{1}$ and $u_{1}$. 

In the second step, we consider the cohomology class of the 
BRST operator $Q_{B}$ acting on ${\pmh}^{(n)}_{Q_{F}}$. 
Denoting the results in the previous subsection, we get 
\begin{eqnarray*}
&&v^{(j,k)}_{1} = Q^{(M-j-k)}_{B} v^{(M-k,M-j)}_{0} \,\,,\\
&&v^{(M-k,M-j)}_{1} = Q^{(j+k)}_{B} v^{(j+M,k+M)}_{0} \,\,,\\
&&v^{(j+M,k+M)}_{0} = Q_{F} \Lambda \hspace{5mm} 
 (\mbox{for some state} \hspace{2mm} \Lambda) \,\,.
\end{eqnarray*}
As a conclusion, we obtain the result, 
\[
{\cal H}^{(m)}_{Q_{B}} {\th}^{(n)}_{Q_{F}} = \left\{
\begin{array}{ll}
{\bv}^{\prime} &  \mbox{ $m=0$ and $n=0$} \\
 0     & \mbox{otherwise}
\end{array}
\right. \,\,,
\]
\begin{eqnarray*}
&&{\cal H}^{(m)}_{Q_{B}} {\hh}^{(n)}_{Q_{F}} =0 \hspace{5mm} \mbox{for all} 
 \hspace{2mm} m \geq 0  \mbox{\hspace{2mm} and \hspace{2mm}} n \geq 0 \,\,,\\
&&{\bv}^{\prime} \equiv \left\{ { \mbox{states in} \hspace{1.5mm} 
 {\FOCK}_{j,k} \hspace{2mm} \mbox{generated by} \hspace{1.5mm}
 \{{v_{0} ,v_{2} ,v_{4} ,\cdots } \} } \right\} \,\,.
\end{eqnarray*}
The contributions of $v_{2i} \hspace{3mm} (i=1,2,3, \cdots)$ seem not to be  
related in the N=2 superconformal module of the Fock space. 
Since they are in the $coker \psi $. So we think that the results 
one desires could be obtained as far as the N=2 superconformal 
module is concerned. 
\subsection[BRST analysis]{Character formula}

\pr
As an application of the above section, we rederive the character 
formula of an N=2 superconformal minimal unitary model. 
A character is a generating function of the multiplicities of states with 
specified quantum numbers $(h,q)$ in the N=2 superconformal module. 
It is defined as follows, 
\begin{eqnarray*}
&&ch( \tau , z) \equiv tr q^{L_{0} -\frac{c}{24}} y^{J_{0}} \,\,,\\
&&q \equiv e^{2 \pi i \tau} \,\,,\\
&&y \equiv e^{2 \pi i z} \hspace{5mm} 
( \tau ,z \in {\bc} , \mbox{Im} \tau >0) \,\,.
\end{eqnarray*}
Let us consider the states in the Fock space ${\FOCK}_{j,k}$.
A primary field $v_{0}$ has the following quantum numbers, 
\begin{eqnarray*}
&&h_{j,k} =\frac{1}{M} \left\{ {jk-\frac{1}{4}} \right\} \,\,,\\
&&q_{j,k} =\frac{1}{M} (j-k) \,\,.
\end{eqnarray*}
The generators 
$J_{-n} ,L_{-n} ,G^{+}_{-n+\frac{1}{2}} ,
G^{-}_{-n+\frac{1}{2}} \mbox{\hspace{3mm}} (n=1,2,3, \cdots)$ of 
the N=2 superconformal algebra contribute to the character as, 
\begin{eqnarray*}
&&J_{-n} \rightarrow q^{n} \,\,,\\
&&L_{-n} \rightarrow q^{n} \,\,,\\
&&G^{+}_{-n+\frac{1}{2}} \rightarrow y q^{n-\frac{1}{2}} \,\,,\\
&&G^{-}_{-n+\frac{1}{2}} \rightarrow y^{-1} q^{n-\frac{1}{2}} \,\,.
\end{eqnarray*}
Since $J_{-n}$ and $L_{-n}$ are bosonic operators 
and $G^{\pm}_{-n+\frac{1}{2}}$ are fermionic ones, their total contributions 
are written,
\[ 
\fprod 
\frac{(1+y q^{n-\frac{1}{2}})(1+ y^{-1} q^{n-\frac{1}{2}})}
{{(1- q^{n})}^{2}} \,\,.
\] 
We have to subtract the null states. To begin with, 
the cohomology classes ${\pmh}^{(0)}_{Q_{F}}$ allow us to obtain the relation, 
\begin{eqnarray*}
&&{\cal O} \equiv q^{L_{0} -\frac{c}{24}} y^{J_{0}} \,\,,\\
&&{tr}_{{\pmh}_{Q_{F}}^{(0)}} \cal O \\
&&= {tr}_{\ker Q_{F}^{\left[{0}\right]}} {\cal O} \\
&&= {tr}_{\ft{(0)}{\pm}} {\cal O} - 
  {tr}_{\mbox{Im} Q_{F}^{\left[{0}\right]}} {\cal O} \\
&&= {tr}_{\ft{(0)}{\pm}} {\cal O} - {tr}_{\ft{(1)}{\pm}} {\cal O} 
 + {tr}_{\ker Q_{F}^{\left[{1}\right]}} {\cal O} \\
&&= \fsum{i=0} (-1)^{i} {tr}_{\ft{(i)}{\pm}} {\cal O} \,\,.
\end{eqnarray*}
Denote that 
\begin{eqnarray*}
&&\ft{(i)}{+} = {\FOCK}_{j+Mi,k} \bigoplus {\FOCK}_{j+M(i-1),k+M} \bigoplus 
 \cdots \bigoplus {\FOCK}_{j,k+Mi} \,\,,\\
&&\ft{(i)}{-} = {\FOCK}_{M(i+1)-k,M-j} \bigoplus {\FOCK}_{Mi-k,2M-j} 
 \bigoplus \cdots \bigoplus {\FOCK}_{M-k,M(i+1)-j} \,\,,
\end{eqnarray*}
we rewrite these formulae,
\begin{eqnarray*}
&&{tr}_{{\th}^{(0)}_{Q_{F}}} {\cal O} = \fsum{l=0} \fsum{m=0} (-1)^{l+m} 
 {tr}_{{\FOCK}_{j+Ml,k+Mm}} {\cal O} \,\,,\\
&&{tr}_{{\hh}^{(0)}_{Q_{F}}} {\cal O} = \fsum{l=0} \fsum{m=0} (-1)^{l+m} 
 {tr}_{{\FOCK}_{M(l+1)-k,M(m+1)-j}} {\cal O} \,\,.
\end{eqnarray*}
Furthermore we may further rewrite these,
\begin{eqnarray*}
\lefteqn{{tr}_{{\th}^{(0)}_{Q_{F}}} {\cal O}} \\
&&= \fsum{l=0} \left[ {{tr}_{{\FOCK}_{j+Ml,k+Ml}} {\cal O} 
 +\fsum{m=1} {(-1)}^{m} \left\{{{tr}_{{\FOCK}_{j+Ml,k+Mm+Ml}}} {\cal O} 
 + {tr}_{{\FOCK}_{j+Mm+Ml,k+Ml}} {\cal O} \right\} } \right] \,\,,\\
\lefteqn{{tr}_{{\hh}^{(0)}_{Q_{F}}} {\cal O}} \\
&&= \fsum{l=0} 
 \left[ {{tr}_{{\FOCK}_{M(l+1)-k,M(l+1)-j}} {\cal O} + \fsum{m=1} {(-1)}^{m} 
    \left\{ {{tr}_{{\FOCK}_{M(l+1)-k,M(l+m+1)-j}} {\cal O}
     +{tr}_{{\FOCK}_{M(l+m+1)-k,M(l+1)-j}} {\cal O} } \right\} } \right] \,\,.
\end{eqnarray*}
Thus we obtain the expressions,
\begin{eqnarray*}
&&{tr}_{{\th}^{(0)}_{Q_{F}}} q^{L_{0} -\frac{c}{24}} y^{J_{0}} \\
&&= \fprod \frac{(1+y q^{n-\frac{1}{2}})(1+ y^{-1} q^{n-\frac{1}{2}})}
{{(1- q^{n})}^{2}} \times \fsum{l=0} 
 \biggl[ q^{\frac{1}{M} \left[{(j+Ml)(k+Ml)-\frac{1}{4}}\right]} 
   y^{\frac{1}{M} (j-k)} \\
&& \mbox{} + \fsum{m=1} (-1)^{m} 
    \left\{{ q^{\frac{1}{M} \left[{(j+Ml)(k+M(l+m)) -\frac{1}{4}}\right]}
     y^{\frac{1}{M} (j-k-Mm)} 
      + q^{\frac{1}{M} \left[{(j+M(l+m))(k+Ml)-\frac{1}{4}}\right] } 
    y^{\frac{1}{M} (j-k+Mm)} } \right\}  \biggr] \,\,,\\
&=& \fprod  \frac{(1+y q^{n-\frac{1}{2}})(1+ y^{-1} q^{n-\frac{1}{2}})}
    {{(1- q^{n})}^{2}} 
\times q^{\frac{1}{M} \left({jk-\frac{1}{4}}\right)} y^{\frac{1}{M} (j-k)} 
   \times  \fsum{l=0} q^{M l^{2} + (j+k)l} \\
&&\times \left[{1-\frac{q^{j+Ml} y^{-1}}{1+ q^{j+Ml} y^{-1}}
   -\frac{q^{k+Ml} y }{1+ q^{k+Ml} y } }\right] \,\,.
\end{eqnarray*}
Similarly we obtain the relation,
\begin{eqnarray*}
\lefteqn{{tr}_{{\hh}^{(0)}_{Q_{F}}} q^{L_{0} -\frac{c}{24}} y^{J_{0}}} \\
&&= \fprod \frac{(1+y q^{n-\frac{1}{2}})(1+ y^{-1} q^{n-\frac{1}{2}})}
{{(1- q^{n})}^{2}} 
\times q^{\frac{1}{M} \left({jk-\frac{1}{4}}\right)} y^{\frac{1}{M} (j-k)}
 \fsum{l=1} q^{M l^{2} -(j+k)l } \\
&&\times 
\left[ {1-\frac{q^{Ml-j} y}{1+ q^{Ml-j} y} 
-\frac{q^{Ml-k} y^{-1}}{1+ q^{Ml-k} y^{-1}} } \right] \,\,.
\end{eqnarray*}
As a final result, we can write down the character formula, 
\begin{eqnarray*}
&&{\cal O} \equiv q^{L_{0} -\frac{c}{24}} y^{J_{0}} \,\,,\\
 ch( \tau ,z ) & \equiv & {tr} {\cal O} \\
&=& {tr}_{{\cal H}_{Q_{B}} {\cal H}_{Q_{F}}} {\cal O} \\
%&=& {tr}_{{\th}^{(0)}_{Q_{F}}} {\cal O} - {tr}_{{\hh}^{(0)}_{Q_{F}}} {\cal O} 
% + {tr}_{{\th}^{(1)}_{Q_{F}}} {\cal O} 
% - {tr}_{{\hh}^{(1)}_{Q_{F}}} {\cal O} +- \cdots \\
%&=& \fsum{l=0} 
%\left[ {{tr}_{{\th}^{(l)}_{Q_{F}}} {\cal O} 
%  -{tr}_{{\hh}^{(l)}_{Q_{F}}} {\cal O} } \right] \\
&=& {tr}_{{\th}^{(0)}_{Q_{F}}} {\cal O} -{tr}_{{\hh}^{(0)}_{Q_{F}}} {\cal O} 
 \mbox{\hspace{33mm}} 
 \left( { {tr}_{{\pmh}^{(l)}_{Q_{F}}} {\cal O} =0  \hspace{2mm} 
 \mbox{for all} \hspace{2mm} l \geq 1} 
   \right) \\
&=& \fprod \frac{(1+y q^{n-\frac{1}{2}})(1+ y^{-1} q^{n-\frac{1}{2}})}
{{(1- q^{n})}^{2}}  \times 
q^{\frac{1}{M} \left({jk-\frac{1}{4}}\right)} y^{\frac{1}{M} (j-k)} \\
&&\times 
 \biggl[ \fsum{l=0} q^{M l^{2} +(j+k)l} 
  \left( { 1-\frac{y^{-1} q^{Ml+j}}{1+ y^{-1} q^{Ml+j}} 
  -\frac{y q^{Ml+k}}{1+y q^{Ml+k}} } \right) \\
&& -\fsum{l=1} q^{M l^{2} -(j+k)l} 
    \left( { 1 -\frac{y q^{Ml-j}}{1+y q^{Ml-j}} 
     -\frac{y^{-1} q^{Ml-k}}{1+ y^{-1} q^{Ml-k} }} \right)  \biggr] \,\,.
\end{eqnarray*}
This formula coincides with the known character formula derived in 
\cite{Matsu,Dob}, and may also be 
re-expressed as an infinite product, 
\begin{eqnarray*}
&& ch( \tau ,z ) \\
&&= \fprod \frac{(1+y q^{n-\frac{1}{2}})(1+ y^{-1} q^{n-\frac{1}{2}})}
{{(1- q^{n})}^{2}}  \times 
q^{\frac{1}{M} \left({jk-\frac{1}{4}}\right)} y^{\frac{1}{M} (j-k)} \\
&&\times
\fprod \frac{(1-q^{Mn+j+k-M})(1- q^{Mn-j-k}){(1- q^{Mn})}^{2}}
{(1+y q^{Mn-j})(1+ y^{-1} q^{Mn+j-M})(1+ y^{-1} q^{Mn-k})(1+y q^{Mn+k-M})} \,\,.
\end{eqnarray*}

\section{Conclusion}

\pr
In this paper, we have analyzed the N=2 superconformal minimal unitary models 
using the BRST cohomology approach. 
The study of the nilpotency property of the 
N=2 bosonic BRST operator has not been done before and  
we think that the research in 
this paper is the first to investigate the analysis 
of N=2 superconformal algebra using the Coulomb gas formalism. 
We investigated the Fock space with $j > 0$ and $k > 0$. 
But the Fock space with $j < 0$ and $k < 0$ 
is isomorphic to the dual Fock space and we did not have to  
consider this case. 
In fact the BRST operators act on the Fock space with $j < 0$ and $k < 0$ 
in the way that the directions of all arrows in the figures 
of the Fock space with $j > 0$ and $k > 0$ 
should be reversed at the same time and exchange the $Q_{f}$ 
and ${\bar{Q}}_{f}$. \\
There are several problems which we did not deal in this paper. \\

1   Correlation function \\
The various correlation functions have already be 
calculated in other contexts, 
but no one deals with them by using a BRST analysis yet. \\

2   Twisted sector \\
In this paper, we considered in the Neveu-Schwarz sector. 
By the spectral flow, this analysis is also 
valid in the Ramond sector. 
However we have not investigated the Fock space in the twisted sector yet 
(It seems to be easier to analyze the twisted sector than the 
Neveu-Schwarz sector).\\

3   Serre relation \\
It might be possible that a Serre relation, if exists, would make 
the analysis in this paper clearer and the intertwiner for some quantum group 
would facilitate to the calculation of the correlation functions.  \\

4  Embedding diagram \\
The embedding diagram which we suggest should be analyzed more 
rigorously by using a filtration technique \cite{Fel}. \\

\vspace{0.2in}
\section{Acknowlegements}

\pr
The author would like to thank Prof.T. Eguchi for suggesting this problem 
and discussions and improving the manuscript.
He also thanks T. Nakatsu and J. Shiraishi for helpful discussions  
and S. Mori for perpetual encouragement. 

\newpage 
\section*{Appendix}
\appendix

\section{Null states} 

\pr
We calculate several null states in the oscillator formalism. \\

$\bullet$ level $1/2$  ( relative charge $+1$ ) ($h= \frac{1}{2} q$) 
\begin{eqnarray}
G^{+}_{-1/2} \aket = -i \bar{\alpha} {b}_{-1/2} \aket \label{eqn:A1} 
\end{eqnarray}

$\bullet$ level $1/2$  ( relative charge $-1$ ) ($h= -\frac{1}{2} q$)
\begin{eqnarray}
G^{-}_{-1/2} \aket = i \alpha {\mb}_{-1/2} \aket \label{eqn:A2} 
\end{eqnarray}

$\bullet$ level $1$ (relative charge $0$)
($h= \frac{q^{2} -\tilde{c}}{2(\tilde{c} -1)}$)
\begin{eqnarray}
&&\left\{ {L_{-1} -\frac{2h+1}{q-1} J_{-1} +\frac{2}{q-1} 
 G^{+}_{-1/2} G^{-}_{-1/2} } \right\} \aket \nonumber  \\
&& =\frac{\beta (\alpha +\bar{\alpha}) +1}{\beta (\alpha -\bar{\alpha})-1} 
 \left[{(\beta + \alpha) {\ma}_{-1} - (\beta +\bar{\alpha}) {a}_{-1}
 - {b}_{-1/2} {\mb}_{-1/2} } \right] \aket \label{eqn:A3}
\end{eqnarray}

$\bullet$ level $3/2$  ( relative charge $+1$ ) 
$(h=\frac{3}{2} q+1- \tilde{c} )$ 
\begin{eqnarray}
&&\left\{ {G^{+}_{-3/2} - \frac{1}{h-\frac{1}{2} q} L_{-1} G^{+}_{-1/2} 
 +\frac{1}{h- \frac{1}{2} q} J_{-1} G^{+}_{-1/2}} \right\} \aket \nonumber \\
&=& i (\bar{\alpha} -\beta) 
 \left\{ {-{b}_{-3/2} +\frac{1}{\alpha + \beta} 
 {a}_{-1} {b}_{-1/2} } \right\} \aket  \label{eqn:A4}
\end{eqnarray}

$\bullet$ level $3/2$  ( relative charge $-1$ ) 
$(h=-\frac{3}{2} q+1- \tilde{c} )$ 
\begin{eqnarray}
&&\left\{ {G^{-}_{-3/2} -\frac{1}{h+\frac{1}{2} q} L_{-1} G^{-}_{-1/2} 
 -\frac{1}{h+\frac{1}{2} q} J_{-1} G^{-}_{-1/2} } \right\} \aket \nonumber \\
&=& i (\alpha -\beta) 
\left\{ {{\mb}_{-3/2} -\frac{1}{{\bar{\alpha}} +\beta} 
 {\ma}_{-1} {\mb}_{-1/2}} \right\} \aket \label{eqn:A5} 
\end{eqnarray}

\newpage

\section*{Figure captions}

\vspace*{4mm}
\noindent
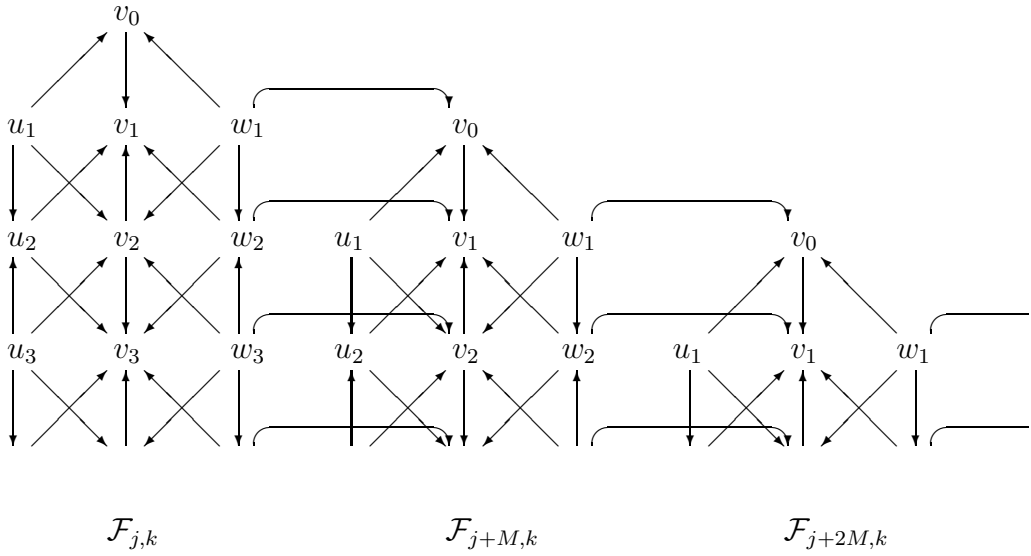
{\bf Fig.1}$\;$
The action of the fermionic BRST operator ${\bar{Q}}_{f}$. This operator maps 
a state in a Fock space ${\cal F}_{j+lM,k}$ \hspace{2mm} $(l=0,1,2, \cdots)$
to a state in the Fock space ${\cal F}_{j+(l+1)M,k}$ \hspace{2mm} 
$(l=0,1,2, \cdots)$ when the image is non-vanishing. 
Actions of the BRST operator ${\bar{Q}}_{f}$ on the states which have 
non-vanishing images are represented by the horizontal oval lines. 
The other states vanish by the actions of the operator 
${\bar{Q}}_{f}$.

\

\noindent
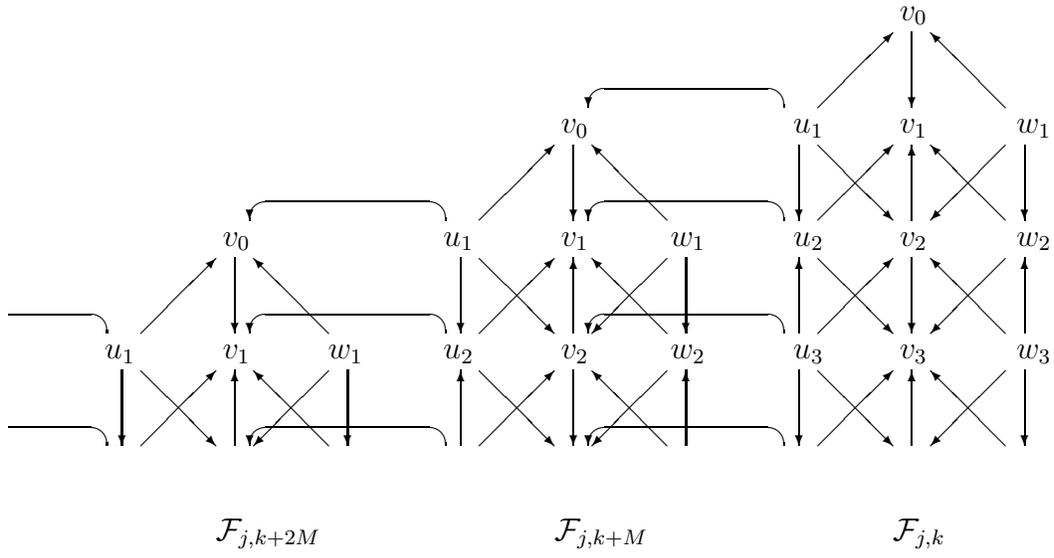
{\bf Fig.2}$\;$
The actions of the fermionic BRST operator $Q_{f}$. This operator maps 
a state in a Fock space ${\cal F}_{j,k+lM}$ \hspace{2mm} $(l=0,1,2, \cdots)$
to a state in the Fock space ${\cal F}_{j,k+(l+1)M}$ \hspace{2mm}
$(l=0,1,2, \cdots)$ when the image is non-vanishing.
Actions of the BRST operator ${Q}_{f}$ on the states which have non-vanishing 
images are represented by the horizontal oval lines. 
The other states vanish by the actions of the operator ${Q}_{f}$. 

\

\noindent
{\bf Fig.3}$\;$
The actions of the bosonic BRST operator ${Q}_{B}$. This operator maps 
a state in a Fock space ${\cal F}_{j+(l+1)M,k+(l+1)M}$ \hspace{1mm}  
$({\cal F}_{(l+1)M-k,(l+1)M-j})$ \hspace{2mm} $(l=0,1,2, \cdots)$ 
to a state in the Fock space ${\cal F}_{(l+1)M-k,(l+1)M-j}$ \hspace{1mm} 
$({\cal F}_{j+lM,k+lM})$ \hspace{2mm} $(l=0,1,2, \cdots)$ 
when the image is non-vanishing. Actions of the BRST operator ${Q}_{B}$
on the states which have non-vanishing images are represented by the 
horizontal oval lines. 
The other states vanish by the actions of the operator ${Q}_{f}$. 

\

\noindent
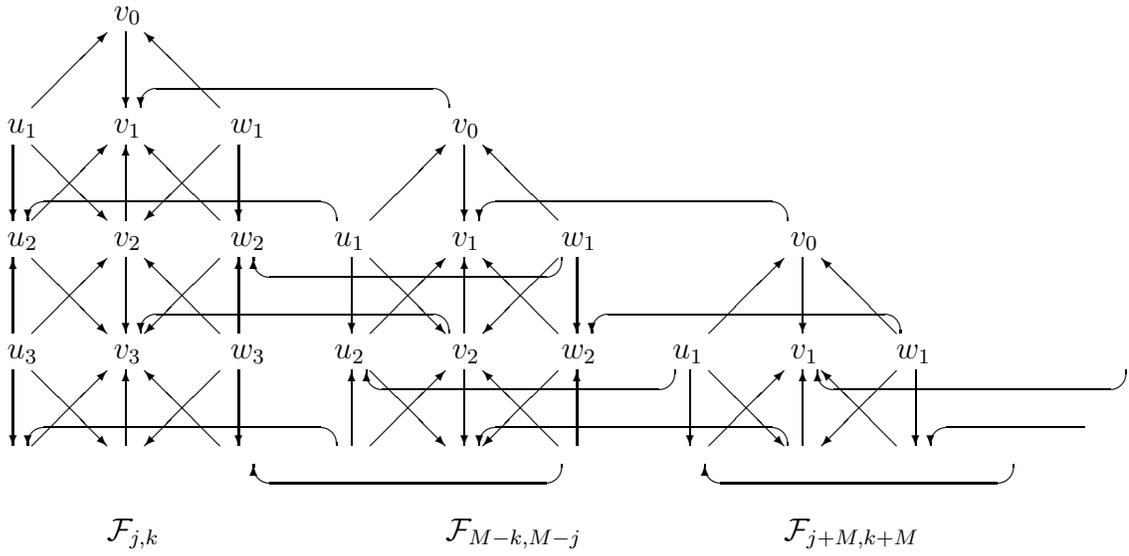
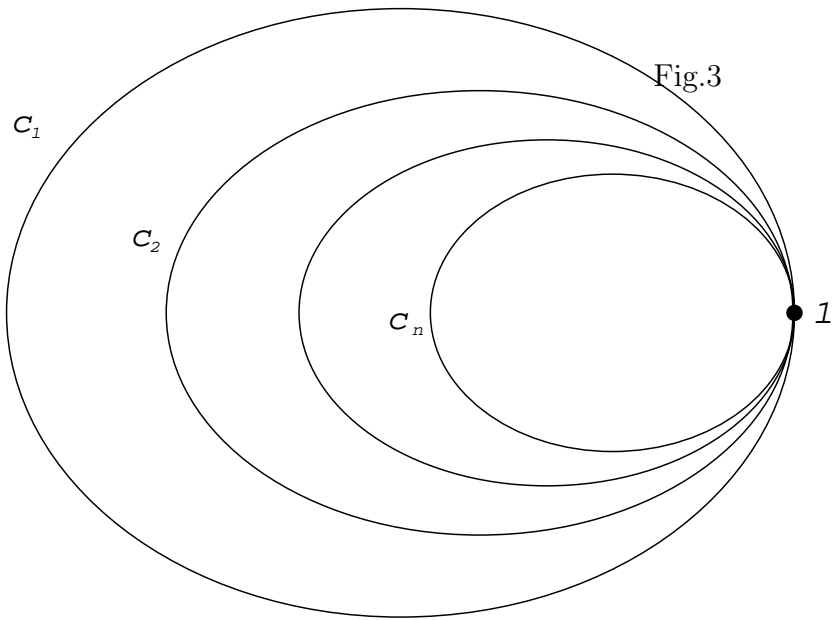
{\bf Fig.4}$\;$
The integral contour of the bosonic BRST operator $Q^{(n)}_{B}$ 
(eq. (\ref{eqn:QB})). The vertex operators $V_{B} (z_{i})$ are arranged 
such that the arguments of the operators satisfy a radial ordering 
($|z_{1}| > |z_{2}| > \cdots > |z_{n}|$). Each integral contour $C_{i}$
is associated with the integral variable $z_{i}$.

\

\noindent
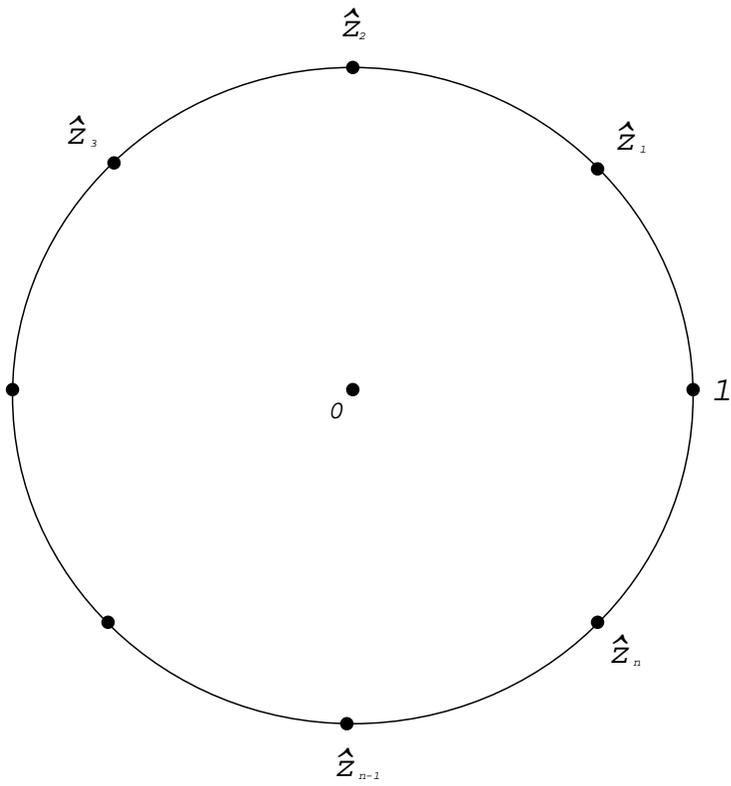
{\bf Fig.5}$\;$
The integral contour of the bosonic BRST operator $Q^{(n)}_{B}$ 
(eq. (\ref{eqn:BQ})). The vertex operators $V_{B} (z_{i})$ are
arranged such that the arguments of these operators satisfy the 
sequence ($0 < \mbox{arg} {z}_{1} < \mbox{arg} {z}_{2} < \cdots <
  \mbox{arg} {z}_{n} < 2 \pi$).
The variables $\hat{{z}_{i}} \equiv {{z}_{i}}/{|{{z}_{i}}|}$ associated
with $V_{B} ({z}_{i})$ $(i=1,2, \cdots ,n)$ are arranged 
on the unit circle counterclockwise.

\newpage
%BRST \bar{Q}_{f}
\unitlength 5mm
\begin{picture}(31,16)
%Fock space F_{j,k}
%\put(4.5,0){${\cal F}_{j,k}$}
\multiput(2,4.5)(0,6){2}{\vector(0,-1){2}}
\multiput(8,4.5)(0,6){2}{\vector(0,-1){2}}
\multiput(5,7.5)(0,6){2}{\vector(0,-1){2}}
\multiput(2,5.5)(6,0){2}{\vector(0,1){2}}
\multiput(5,2.5)(0,6){2}{\vector(0,1){2}}
\multiput(2.5,2.5)(0,3){4}{\vector(1,1){2}}
\multiput(2.5,4.5)(0,3){3}{\vector(1,-1){2}}
\multiput(7.5,2.5)(0,3){4}{\vector(-1,1){2}}
\multiput(7.5,4.5)(0,3){3}{\vector(-1,-1){2}}
\put(2,10.7){\makebox(0.5,0.5){${u}_{1}$}}
\put(2,7.7){\makebox(0.5,0.5){${u}_{2}$}}
\put(2,4.7){\makebox(0.5,0.5){${u}_{3}$}}
\put(4.8,13.7){\makebox(0.5,0.5){${v}_{0}$}}
\put(4.8,10.7){\makebox(0.5,0.5){${v}_{1}$}}
\put(4.8,7.7){\makebox(0.5,0.5){${v}_{2}$}}
\put(4.8,4.7){\makebox(0.5,0.5){${v}_{3}$}}
\put(8,10.7){\makebox(0.5,0.5){${w}_{1}$}}
\put(8,7.7){\makebox(0.5,0.5){${w}_{2}$}}
\put(8,4.7){\makebox(0.5,0.5){${w}_{3}$}}

%Fock space F_{j+M,k}
%\put(13.5,0){${\cal F}_{j+M,k}$}
\multiput(11,2.5)(6,0){2}{\vector(0,1){2}}
\multiput(14,4.5)(0,6){2}{\vector(0,-1){2}}
\multiput(11,7.5)(6,0){2}{\vector(0,-1){2}}
\multiput(11.5,2.5)(0,3){3}{\vector(1,1){2}}
\multiput(11.5,4.5)(0,3){2}{\vector(1,-1){2}}
\multiput(16.5,2.5)(0,3){3}{\vector(-1,1){2}}
\multiput(16.5,4.5)(0,3){2}{\vector(-1,-1){2}}
\put(14,5.5){\vector(0,1){2}}
\put(10.7,7.7){\makebox(0.5,0.5){$u_{1}$}}
\put(10.7,4.7){\makebox(0.5,0.5){$u_{2}$}}
\put(13.8,10.7){\makebox(0.5,0.5){$v_{0}$}}
\put(13.8,7.7){\makebox(0.5,0.5){$v_{1}$}}
\put(13.8,4.7){\makebox(0.5,0.5){$v_{2}$}}
\put(16.8,7.7){\makebox(0.5,0.5){$w_{1}$}}
\put(16.8,4.7){\makebox(0.5,0.5){$w_{2}$}}

%Fock space F_{j+2M,k}
%\put(22.5,0){${\cal F}_{j+2M,k}$}
\multiput(20,4.5)(6,0){2}{\vector(0,-1){2}}
\multiput(20.5,2.5)(0,3){2}{\vector(1,1){2}}
\multiput(25.5,2.5)(0,3){2}{\vector(-1,1){2}}
\put(23,2.5){\vector(0,1){2}}
\put(23,7.5){\vector(0,-1){2}}
\put(20.5,4.5){\vector(1,-1){2}}
\put(25.5,4.5){\vector(-1,-1){2}}
\put(19.7,4.7){\makebox(0.5,0.5){$u_{1}$}}
\put(22.8,7.7){\makebox(0.5,0.5){$v_{0}$}}
\put(22.8,4.7){\makebox(0.5,0.5){$v_{1}$}}
\put(25.7,4.7){\makebox(0.5,0.5){$w_{1}$}}
\multiput(11,2.5)(0,3){4}{\oval(5.2,1)[t]}
\multiput(20,2.5)(0,3){3}{\oval(5.2,1)[t]}
\multiput(29,2.5)(0,3){2}{\oval(5.2,1)[tl]}
\multiput(13.6,2.5)(0,3){4}{\vector(0,-1){0}}
\multiput(22.6,2.5)(0,3){3}{\vector(0,-1){0}}
\put(4.5,0){${\cal F}_{j,k}$}
\put(13.5,0){${\cal F}_{j+M,k}$}
\put(22.5,0){${\cal F}_{j+2M,k}$}
\put(16,-4){Fig.1}
\end{picture}

\newpage
%BRST Q_{f}
\unitlength 5mm
\begin{picture}(31,16)
%Fock space F_{j,k+2M}
%\put(4.5,0){${\cal F}_{j,k+2M}$}
\multiput(20,4.5)(0,6){2}{\vector(0,-1){2}}
\multiput(26,4.5)(0,6){2}{\vector(0,-1){2}}
\multiput(23,7.5)(0,6){2}{\vector(0,-1){2}}
\multiput(20,5.5)(6,0){2}{\vector(0,1){2}}
\multiput(23,2.5)(0,6){2}{\vector(0,1){2}}
\multiput(20.5,2.5)(0,3){4}{\vector(1,1){2}}
\multiput(20.5,4.5)(0,3){3}{\vector(1,-1){2}}
\multiput(25.5,2.5)(0,3){4}{\vector(-1,1){2}}
\multiput(25.5,4.5)(0,3){3}{\vector(-1,-1){2}}
\put(20,10.7){\makebox(0.5,0.5){${u}_{1}$}}
\put(20,7.7){\makebox(0.5,0.5){${u}_{2}$}}
\put(20,4.7){\makebox(0.5,0.5){${u}_{3}$}}
\put(22.8,13.7){\makebox(0.5,0.5){${v}_{0}$}}
\put(22.8,10.7){\makebox(0.5,0.5){${v}_{1}$}}
\put(22.8,7.7){\makebox(0.5,0.5){${v}_{2}$}}
\put(22.8,4.7){\makebox(0.5,0.5){${v}_{3}$}}
\put(26,10.7){\makebox(0.5,0.5){${w}_{1}$}}
\put(26,7.7){\makebox(0.5,0.5){${w}_{2}$}}
\put(26,4.7){\makebox(0.5,0.5){${w}_{3}$}}

%Fock space F_{j,k+M}
%\put(13.5,0){${\cal F}_{j,k+M}$}
\multiput(11,2.5)(6,0){2}{\vector(0,1){2}}
\multiput(14,4.5)(0,6){2}{\vector(0,-1){2}}
\multiput(11,7.5)(6,0){2}{\vector(0,-1){2}}
\multiput(11.5,2.5)(0,3){3}{\vector(1,1){2}}
\multiput(11.5,4.5)(0,3){2}{\vector(1,-1){2}}
\multiput(16.5,2.5)(0,3){3}{\vector(-1,1){2}}
\multiput(16.5,4.5)(0,3){2}{\vector(-1,-1){2}}
\put(14,5.5){\vector(0,1){2}}
\put(10.7,7.7){\makebox(0.5,0.5){$u_{1}$}}
\put(10.7,4.7){\makebox(0.5,0.5){$u_{2}$}}
\put(13.8,10.7){\makebox(0.5,0.5){$v_{0}$}}
\put(13.8,7.7){\makebox(0.5,0.5){$v_{1}$}}
\put(13.8,4.7){\makebox(0.5,0.5){$v_{2}$}}
\put(16.8,7.7){\makebox(0.5,0.5){$w_{1}$}}
\put(16.8,4.7){\makebox(0.5,0.5){$w_{2}$}}

%Fock space F_{j,k}
%\put(22.5,0){${\cal F}_{j,k}$}
\multiput(2,4.5)(6,0){2}{\vector(0,-1){2}}
\multiput(2.5,2.5)(0,3){2}{\vector(1,1){2}}
\multiput(7.5,2.5)(0,3){2}{\vector(-1,1){2}}
\put(5,2.5){\vector(0,1){2}}
\put(5,7.5){\vector(0,-1){2}}
\put(2.5,4.5){\vector(1,-1){2}}
\put(7.5,4.5){\vector(-1,-1){2}}
\put(1.7,4.7){\makebox(0.5,0.5){$u_{1}$}}
\put(4.8,7.7){\makebox(0.5,0.5){$v_{0}$}}
\put(4.8,4.7){\makebox(0.5,0.5){$v_{1}$}}
\put(7.7,4.7){\makebox(0.5,0.5){$w_{1}$}}
\multiput(-1,2.5)(0,3){2}{\oval(5.2,1)[tr]}
\multiput(8,2.5)(0,3){3}{\oval(5.2,1)[t]}
\multiput(17,2.5)(0,3){4}{\oval(5.2,1)[t]}
\multiput(5.4,2.5)(0,3){3}{\vector(0,-1){0}}
\multiput(14.4,2.5)(0,3){4}{\vector(0,-1){0}}
\put(4.5,0){${\cal F}_{j,k+2M}$}
\put(13.5,0){${\cal F}_{j,k+M}$}
\put(22.5,0){${\cal F}_{j,k}$}
\put(16,-4){Fig.2}
\end{picture}

\newpage
%BRST Q^{+}
\unitlength 5mm
\begin{picture}(31,16)
%Fock space F_{j,k}
%\put(4.5,0){${\cal F}_{j,k}$}
\multiput(2,4.5)(0,6){2}{\vector(0,-1){2}}
\multiput(8,4.5)(0,6){2}{\vector(0,-1){2}}
\multiput(5,7.5)(0,6){2}{\vector(0,-1){2}}
\multiput(2,5.5)(6,0){2}{\vector(0,1){2}}
\multiput(5,2.5)(0,6){2}{\vector(0,1){2}}
\multiput(2.5,2.5)(0,3){4}{\vector(1,1){2}}
\multiput(2.5,4.5)(0,3){3}{\vector(1,-1){2}}
\multiput(7.5,2.5)(0,3){4}{\vector(-1,1){2}}
\multiput(7.5,4.5)(0,3){3}{\vector(-1,-1){2}}
\put(2,10.7){\makebox(0.5,0.5){${u}_{1}$}}
\put(2,7.7){\makebox(0.5,0.5){${u}_{2}$}}
\put(2,4.7){\makebox(0.5,0.5){${u}_{3}$}}
\put(4.8,13.7){\makebox(0.5,0.5){${v}_{0}$}}
\put(4.8,10.7){\makebox(0.5,0.5){${v}_{1}$}}
\put(4.8,7.7){\makebox(0.5,0.5){${v}_{2}$}}
\put(4.8,4.7){\makebox(0.5,0.5){${v}_{3}$}}
\put(8,10.7){\makebox(0.5,0.5){${w}_{1}$}}
\put(8,7.7){\makebox(0.5,0.5){${w}_{2}$}}
\put(8,4.7){\makebox(0.5,0.5){${w}_{3}$}}

%Fock space F_{M-k,M-j}
%\put(13.5,0){${\cal F}_{M-k,M-j}$}
\multiput(11,2.5)(6,0){2}{\vector(0,1){2}}
\multiput(14,4.5)(0,6){2}{\vector(0,-1){2}}
\multiput(11,7.5)(6,0){2}{\vector(0,-1){2}}
\multiput(11.5,2.5)(0,3){3}{\vector(1,1){2}}
\multiput(11.5,4.5)(0,3){2}{\vector(1,-1){2}}
\multiput(16.5,2.5)(0,3){3}{\vector(-1,1){2}}
\multiput(16.5,4.5)(0,3){2}{\vector(-1,-1){2}}
\put(14,5.5){\vector(0,1){2}}
\put(10.7,7.7){\makebox(0.5,0.5){$u_{1}$}}
\put(10.7,4.7){\makebox(0.5,0.5){$u_{2}$}}
\put(13.8,10.7){\makebox(0.5,0.5){$v_{0}$}}
\put(13.8,7.7){\makebox(0.5,0.5){$v_{1}$}}
\put(13.8,4.7){\makebox(0.5,0.5){$v_{2}$}}
\put(16.8,7.7){\makebox(0.5,0.5){$w_{1}$}}
\put(16.8,4.7){\makebox(0.5,0.5){$w_{2}$}}

%Fock space F_{j+M,k+M}
%\put(22.5,0){${\cal F}_{j+M,k+M}$}
\multiput(20,4.5)(6,0){2}{\vector(0,-1){2}}
\multiput(20.5,2.5)(0,3){2}{\vector(1,1){2}}
\multiput(25.5,2.5)(0,3){2}{\vector(-1,1){2}}
\put(23,2.5){\vector(0,1){2}}
\put(23,7.5){\vector(0,-1){2}}
\put(20.5,4.5){\vector(1,-1){2}}
\put(25.5,4.5){\vector(-1,-1){2}}
\put(19.7,4.7){\makebox(0.5,0.5){$u_{1}$}}
\put(22.8,7.7){\makebox(0.5,0.5){$v_{0}$}}
\put(22.8,4.7){\makebox(0.5,0.5){$v_{1}$}}
\put(25.7,4.7){\makebox(0.5,0.5){$w_{1}$}}
\multiput(6.5,2.5)(0,6){2}{\oval(8.2,1)[t]}
\multiput(18.5,2.5)(0,6){2}{\oval(8.2,1)[t]}
\multiput(9.5,5.5)(0,6){2}{\oval(8.2,1)[t]}
\put(12.5,2){\oval(8.2,1)[b]}
\put(12.5,7.5){\oval(8.2,1)[b]}
\put(15.5,4.5){\oval(8.2,1)[b]}
\put(21.5,5.5){\oval(8.2,1)[t]}
\put(24.5,2){\oval(8.2,1)[b]}
\put(27.5,4.5){\oval(8.2,1)[b]}
\put(30.5,2.5){\oval(8.2,1)[tl]}
\multiput(2.4,2.5)(0,6){2}{\vector(0,-1){0}}
\multiput(14.4,2.5)(0,6){2}{\vector(0,-1){0}}
\multiput(5.4,5.5)(0,6){2}{\vector(0,-1){0}}
\put(8.4,2){\vector(0,1){0}}
\put(8.4,7.5){\vector(0,1){0}}
\multiput(11.4,4.5)(12,0){2}{\vector(0,1){0}}
\put(17.4,5.5){\vector(0,-1){0}}
\put(20.4,2){\vector(0,1){0}}
\put(26.4,2.5){\vector(0,-1){0}}
\put(4.5,0){${\cal F}_{j,k}$}
\put(13.5,0){${\cal F}_{M-k,M-j}$}
\put(22.5,0){${\cal F}_{j+M,k+M}$}
\put(16,-4){Fig.3}
\end{picture}

\begin{figure}[htbp]
\begin{center}
\epsfile{file=fig4.eps}
\end{center}
\end{figure}

\newpage
\begin{figure}[htbp]
\begin{center}
\epsfile{file=fig5.eps}
\end{center}
\end{figure}

\end{document}